\pdfoutput=1

\documentclass[onecolumn,epjc3]{svjour3}

\RequirePackage[T1]{fontenc}

\smartqed

\RequirePackage{graphicx}
\RequirePackage{amssymb}
\RequirePackage{epsf}
\RequirePackage{psfig}
\RequirePackage{mathptmx}
\RequirePackage{flushend}
\RequirePackage{natbib}
\RequirePackage{hyperref}

\journalname{Celestial Mechanics and Dynamical Astronomy}

\begin{document}

\title{Crash test for the Copenhagen problem with oblateness}

\author{Euaggelos E. Zotos}

\institute{Department of Physics, School of Science, \\
Aristotle University of Thessaloniki, \\
GR-541 24, Thessaloniki, Greece \\
Corresponding author's email: {evzotos@physics.auth.gr}}

\date{Received: 4 October 2014 / Revised: 4 February 2015 / Accepted: 17 February 2015 / Published online: 5 March 2015}

\titlerunning{Crash test for the restricted three body problem with oblateness}

\authorrunning{Euaggelos E. Zotos}

\maketitle

\begin{abstract}

The case of the planar circular restricted three-body problem where one of the two primaries is an oblate spheroid is investigated. We conduct a thorough numerical analysis on the phase space mixing by classifying initial conditions of orbits and distinguishing between three types of motion: (i) bounded, (ii) escape and (iii) collisional. The presented outcomes reveal the high complexity of this dynamical system. Furthermore, our numerical analysis shows a strong dependence of the properties of the considered escape basins with the total orbital energy, with a remarkable presence of fractal basin boundaries along all the escape regimes. Interpreting the collisional motion as leaking in the phase space we related our results to both chaotic scattering and the theory of leaking Hamiltonian systems. We also determined the escape and collisional basins and computed the corresponding escape/crash times. The highly fractal basin boundaries observed are related with high sensitivity to initial conditions thus implying an uncertainty between escape solutions which evolve to different regions of the phase space. We hope our contribution to be useful for a further understanding of the escape and crash mechanism of orbits in this version of the restricted three-body problem.


\end{abstract}

\section{Introduction}
\label{intro}

The issue of escape in Hamiltonian systems is a classical problem in nonlinear dynamics (e.g., \citet{C90,CK92,CKK93,STN02}). For energy levels above the escape energy the equipotential surfaces are open and exit channels emerge through which the particles can escape to infinity. The literature is replete with studies of such ``open" Hamiltonian systems (e.g., \citet{BBS09,EP14,KSCD99,NH01,Z14a,Z14ip}). Nevertheless, the issue of escaping orbits in Hamiltonian systems is by far less explored than the closely related problem of chaotic scattering. In this situation, a test particle coming from infinity approaches and then scatters off a complex potential. This phenomenon is well investigated as well interpreted from the viewpoint of chaos theory (e.g., \citet{BOG89,BTS96,BST98,JLS99,JMS95,JT91,SASL06,SSL07}). At this point we should emphasize, that all the above-mentioned references on escapes and chaotic scattering in Hamiltonian system are exemplary rather than exhaustive, taking into account that a vast quantity of related literature exists.

In open Hamiltonian systems an issue of paramount importance is the determination of the basins of escape, similar to basins of attraction in dissipative systems or even the Newton-Raphson fractal structures. An escape basin is defined as a local set of initial conditions of orbits for which the test particles escape through a certain exit in the equipotential surface for energies above the escape value. Basins of escape have been studied in many earlier papers (e.g., \citet{BGOB88,KY91,PCOG96}]). The reader can find more details regarding basins of escape in \citep{C02}, while the review \citep{Z14ip} provides information about the escape properties of orbits in a multi-channel dynamical system of a two-dimensional perturbed harmonic oscillator. The boundaries of an escape basins may be fractal (e.g., \citet{AVS09,BGOB88}) or even respect the more restrictive Wada property (e.g., \citet{AVS01}), in the case where three or more escape channels coexist in the equipotential surface. Escaping orbits in the classical Restricted Three-Body Problem (RTBP) is another typical example (e.g., \citealp{N04,N05,dAT14}).

The classical RTBP assumes that the masses of the two primaries are spherically symmetrical in homogeneous layers however, it is found that several celestial bodies, such as Saturn and Jupiter are sufficiently oblate \citep{BPC99}. In addition, the minor planets (e.g., Ceres) and meteoroids have irregular shapes \citep{MWF87,NC08}. The oblateness or triaxiality of a celestial body can produce perturbation deviations from the two-body motion. The most striking example of perturbations arising from oblateness in the solar system is the orbit of the fifth satellite of Jupiter, Amalthea. This planet is so oblate and the satellite's orbit is so small that its line of apsides advances about $900^{\circ}$ in a year (e.g., \citet{M14}). The study of oblateness includes the series of works of \citet{BS12,MRVK00,KMP05,KDP06,KPP08,KGK12,PPK12,SSR79,SSR86,SRS88,SRS97,S81,S87,S89,S90} by considering the more massive primary as an oblate spheroid with its equatorial plane co-incident with the plane of motion of the primaries.

It is well known that all the planets of the Solar System are not perfect spheres but spheroidals. Therefore when someone desires to model a three-body system it is essential to properly modify the RTBP in order to take into account the exact shape of the planets (primary bodies). In \citet{OV03} the authors compare observational data from the systems Saturn-Tethys-satellite and Saturn-Dione-satellite and they conclude that the corresponding theoretical data are much more accurate when the oblateness of Saturn is taken into consideration. In the same vein, in the work of \citet{SYC08} regarding the dynamics of a spacecraft in the Neptune-Triton problem the oblateness coefficient is also included. In the present paper we continue the work initiated in \citet{N04} and \citet{N05} (hereafter Paper I and II) following the same numerical techniques. Our aim is to numerically investigate the properties of motion in the RTBP with oblateness. As far as we know, this is the first detailed and systematic numerical analysis on the phase space mixing of bounded motion, escape and crash in the RTBP with oblateness and this is exactly the novelty and the contribution of the current work.

The paper is organized as follows: In Section \ref{mod} we introduce the considered dynamical model and we present its properties along with some necessary details. All the computational methods we used in order to determine the character of orbits are described in Section \ref{cometh}. In the following Section, we conduct a thorough numerical investigation revealing the overall orbital structure (bounded regions and basins of escape/crash) of system and how it is affected by the total orbital energy considering three cases regarding the value of the oblateness coefficient. In Section \ref{over} a general overview is provided showing in more detail the influence of the energy as well as the oblateness parameter. Our paper ends with Section \ref{disc}, where the discussion and the conclusions of this research are given.

\section{Details of the dynamical model}
\label{mod}

\begin{figure*}[!tH]
\centering
\resizebox{0.6\hsize}{!}{\includegraphics{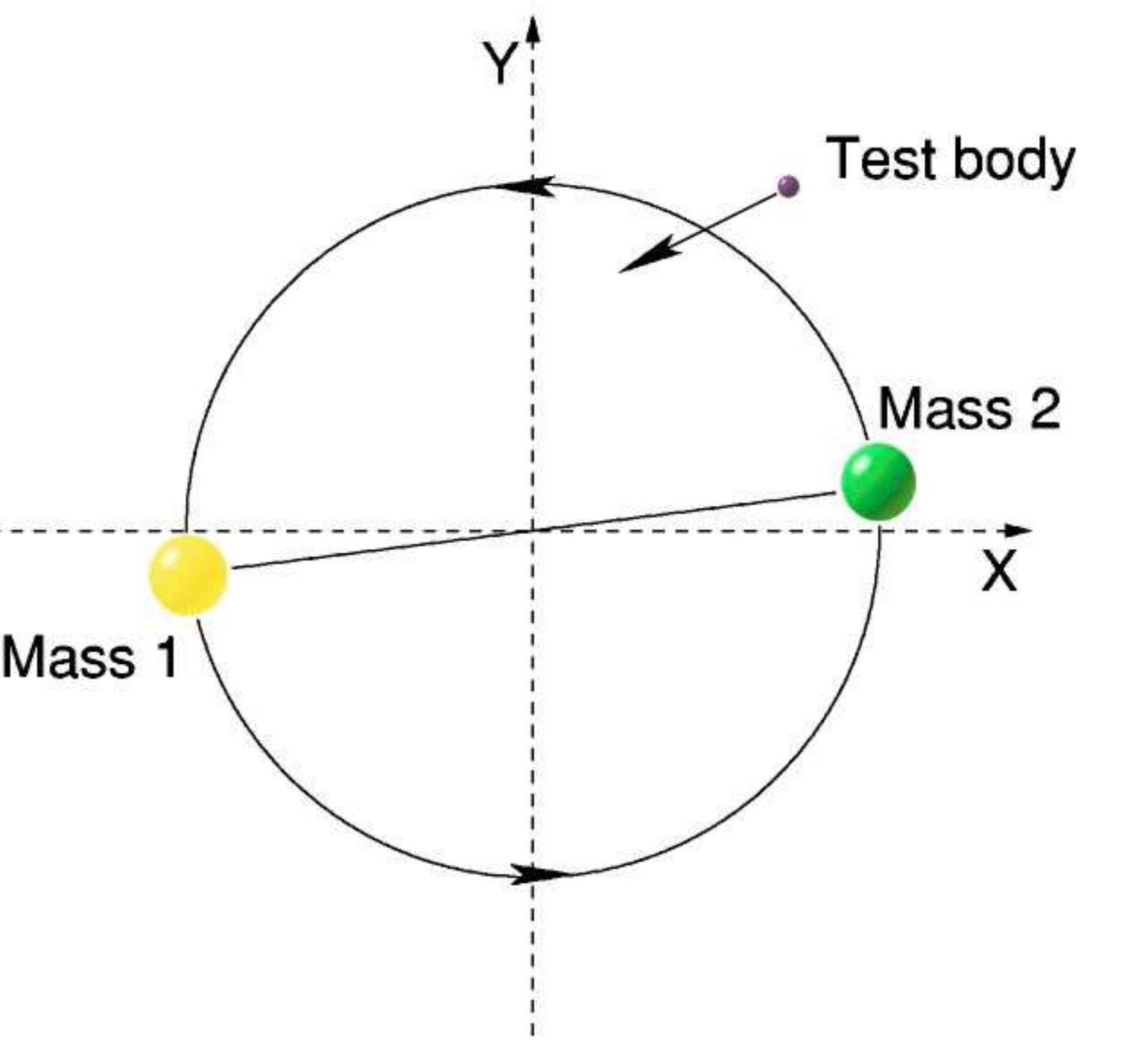}}
\caption{Schematic picture of the planar circular restricted three-body problem for the Copenhagen case.}
\label{rtbp}
\end{figure*}

Let us briefly recall the basic properties and some aspects of the planar restricted three–body problem \citep{S67}. The two primaries move on circular orbits with the same Kepler frequency around their common center of gravity, which is assumed to be fixed at the origin of the coordinates. The third body (test particle with mass much smaller than the masses of the primaries) moves in the same plane under the gravitational field of the two primaries (see Fig. \ref{rtbp}). The non-dimensional masses of the two primaries are $1-\mu$ and $\mu$, where $\mu = m_2/(m_1 + m_2)$ is the mass ratio. We consider the Copenhagen case where $\mu = 1/2$.

We choose as a reference frame a rotating coordinate system where the origin is at (0,0), while the centers $C_1$ and $C_2$ of the two primaries are located at $(-\mu, 0)$ and $(1-\mu,0)$, respectively. The total time-independent gravitational potential is \citep{SSR76}
\begin{equation}
V(x,y) = - \frac{\mu}{r_2} - \frac{(1 - \mu)}{r_1} - \frac{(1 - \mu)A_1}{2r_1^3} - \frac{n^2}{2}\left( x^2  + y^2 \right),
\label{pot}
\end{equation}
where
\begin{equation}
r_1 = \sqrt{\left(x + \mu\right)^2 + y^2},  \ \ \ r_2 = \sqrt{\left(x + \mu - 1\right)^2 + y^2},  \ \ \ n^2 = 1 + \frac{3 A_1}{2},
\label{dist}
\end{equation}
are the distances to the respective primaries and the angular velocity $(n)$, while $A_1$ is the oblateness coefficient which is defined as
\begin{equation}
A_1 = \frac{(RE)^2 - (RP)^2}{5R^2},
\label{obl}
\end{equation}
where $RE$ and $RP$ are the equatorial and polar radius, respectively of the oblate primary, while $R$ is the distance between the centers of the two primaries. We consider values of the oblateness coefficient is in the interval $[0,0.1]$ (see e.g., \citep{KDP06,PK06}), while we study the effect of oblateness up to the linear coefficient $J_2$ only \citep{A12}. We must point that the values of the oblateness in the Solar system are relatively low (e.g, Table 1 in \citet{SSR76}).

The scaled equations of motion describing the motion of the test body in the corotating frame read \citep{SSR76}
\begin{equation}
\ddot{x} = 2n\dot{y} - \frac{\partial V(x,y)}{\partial x},  \ \ \ \ddot{y} = - 2n\dot{x} - \frac{\partial V(x,y)}{\partial y}.
\label{eqmot}
\end{equation}
The dynamical system (\ref{eqmot}) admits the well know Jacobi integral
\begin{equation}
J(x,y,\dot{x},\dot{y}) = \frac{1}{2} \left(\dot{x}^2 + \dot{y}^2 \right) + V(x,y) = E,
\label{ham}
\end{equation}
where $\dot{x}$ and $\dot{y}$ are the momenta per unit mass, conjugate to $x$ and $y$, respectively, while $E$ is the numerical value of the energy which is conserved and defines a three-dimensional invariant manifold in the total four-dimensional phase space. Thus, an orbit with a given value of it's energy integral is restricted in its motion to regions in which $E \leq V(x,y)$, while all other regions are forbidden to the test body. It is widely believed that $J$ is the only independent integral of motion for the PCRTBP system \citep{P93}. The energy value $E$ is related with the Jacobi constant by $C = - 2E$.

\begin{figure*}[!tH]
\centering
\resizebox{0.6\hsize}{!}{\includegraphics{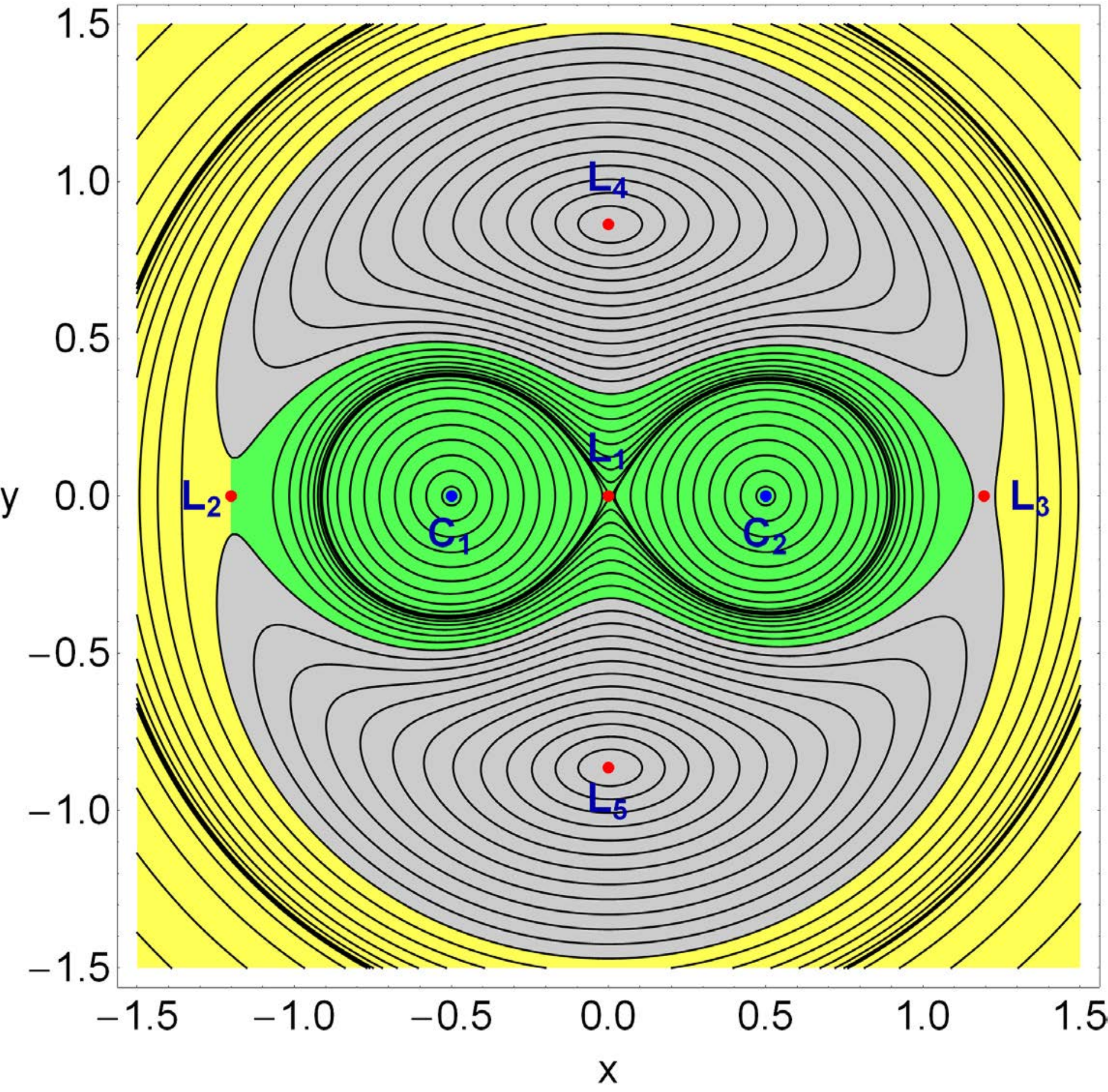}}
\caption{The isolines contours of the constant potential, the location of the centers of the two primaries (blue) and the position of the five Lagrangian points (red), for $A_1 = 0.01$. The interior region is indicated in green, the exterior region is shown in yellow, while the forbidden regions of motion are marked with grey.}
\label{conts}
\end{figure*}

The dynamical system has five equilibria known as Lagrangian points \citep{S67} at which
\begin{equation}
\frac{\partial V(x,y)}{\partial x} = \frac{\partial V(x,y)}{\partial y} = 0.
\label{lps}
\end{equation}
The isolines contours of constant potential, the position of the five Lagrangian points $L_i, \ i = {1,5}$, as well as the centers of the two primaries are shown in Fig. \ref{conts} where $A_1 = 0.01$. Three of them, $L_1$, $L_2$, and $L_3$, are collinear points located in the $x$-axis. We note here that the isolines contours are symmetrical only with respect to the $y = 0$ axis, where the symmetry to the $x = 0$ axis is lost due to the oblateness. The central stationary point $L_1$ is a local minimum of the potential $V(x,y)$. The stationary points $L_2$ and $L_3$ are saddle point. Let $L_2$ located at $x < 0$, while $L_3$ be at $x > 0$. The points $L_4$ and $L_5$ on the other hand, are local maxima of the gravitational potential, enclosed by the banana-shaped isolines. The projection of the four-dimensional phase space onto the physical (or position) space $(x,y)$ is called the Hill's regions and is divided into three domains shown in Fig. \ref{conts} with different colors: (i) the interior region (green) for $x(L_2) \leq x \leq x(L_3)$; (ii) the exterior region (yellow) for $x < x(L_2)$ and $x > x(L_3)$; (iii) the forbidden regions (gray). The boundaries of these Hill's regions are called Zero Velocity Curves (ZVCs) because they are the locus in the physical $(x,y)$ space where the kinetic energy vanishes. Finally, the position of the Lagrangian points $L_2$ and $L_3$ is a function of the oblateness coefficient $A_1$ (e.g., \citet{SL12}).

\begin{figure*}[!tH]
\centering
\resizebox{0.8\hsize}{!}{\includegraphics{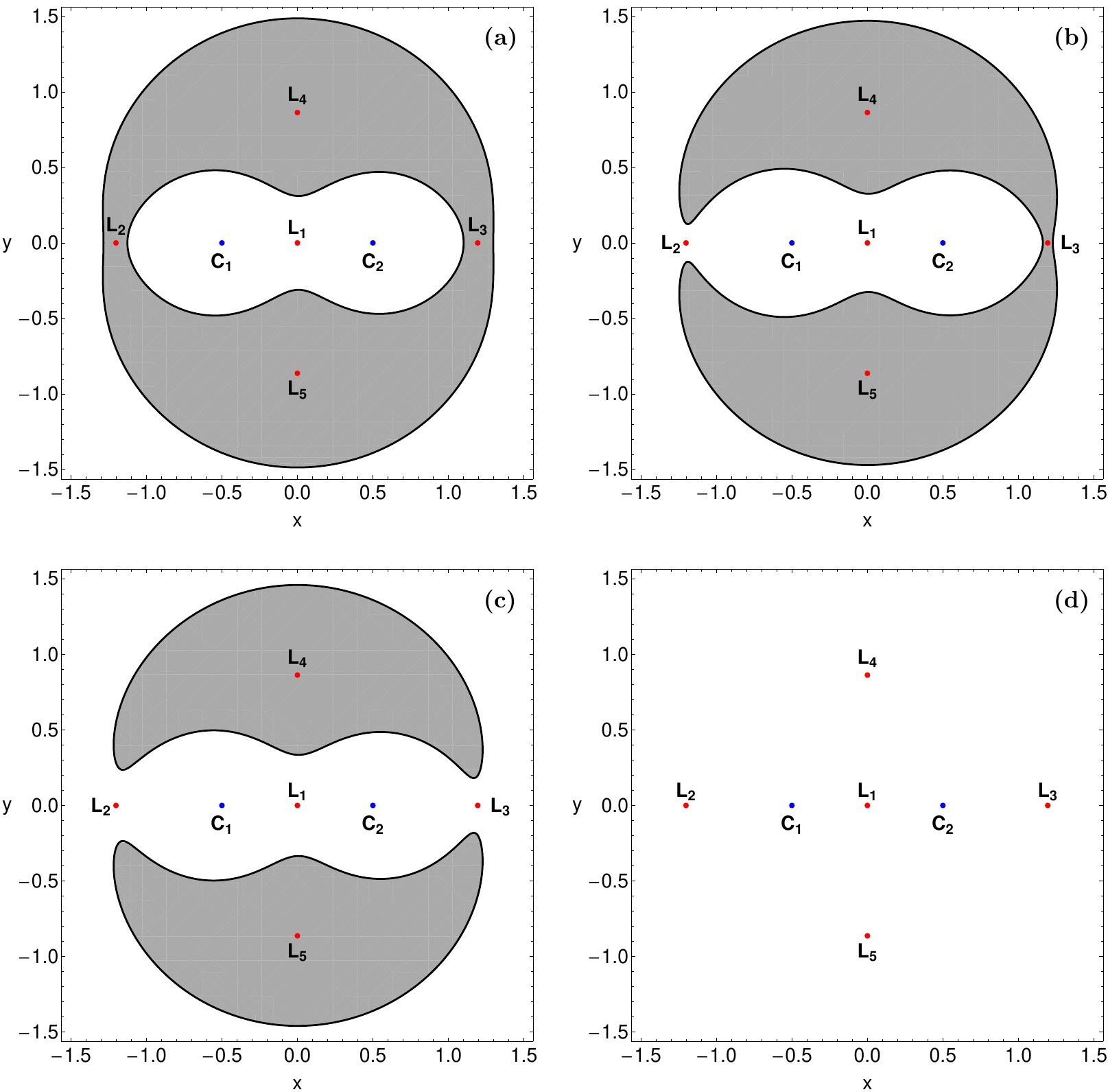}}
\caption{Four possible Hill's region configurations for the PCRTBP system when $A_1 = 0.01$. The white domains correspond to the Hill's region, gray shaded domains indicate the forbidden regions, while the thick black lines depict the Zero Velocity Curves (ZVCs). The red dots pinpoint the position of the Lagrangian points, while the positions of the centers of the two primaries are indicated by blue dots. (a-upper left): $E = -1.76$; (b-upper right): $E = -1.742$; (c-lower left): $E = -1.73$; (d-lower right): $E = -1.30$.}
\label{isos}
\end{figure*}

The values of the Jacobi integral at the five Lagrangian points $L_i$ are critical energy levels and are denoted as $E_i$ (Note that $E_1 = 0$, while $E_4 = E_5$). The structure of the equipotential surfaces strongly depends on the value of the energy. In particular, there are four distinct cases
\begin{itemize}
  \item $E < E_2$: Both necks at $L_2$ and $L_3$ are closed, so in the interior region we have only collisional and bounded motion.
  \item $E_2 < E < E_3$: Only the neck around $L_2$ is open which acts as escape channel.
  \item $E_3 < E < E_4$: The necks around both $L_2$ and $L_3$ are open and two symmetrical, with respect to the $y = 0$ axis, forbidden regions are present.
  \item $E > E_4$: The banana-shaped forbidden regions disappear and therefore, motion over the entire physical $(x,y)$ plane is possible.
\end{itemize}
In Fig. \ref{isos}(a-d) we present for $A_1 = 0.01$ a characteristic equipotential surface for the four possible Hill's region configurations. We observe in Fig. \ref{isos}c the two openings (exit channels) at the Lagrangian points $L_2$ and $L_3$ through which the body can leak out. In fact, we may say that these two exits act as hoses connecting the interior region of the system where $x(L_2) < x < x(L_3)$ with the ``outside world" of the exterior region.

\section{Computational methods and criteria}
\label{cometh}

The motion of the test third body is restricted to a three-dimensional surface $E = const$, due to the existence of the Jacobi integral. With polar coordinates $(r,\phi)$ in the center of the mass system of the corotating frame the condition $\dot{r} = 0$ defines a two-dimensional surface of section, with two disjoint parts $\dot{\phi} < 0$ and $\dot{\phi} > 0$. Each of these two parts has a unique projection onto the configuration physical $(x,y)$ space. Our investigation takes place in both types of projection for a better understanding of the orbital dynamics. In order to explore the behavior of test particles in the Copenhagen model, we need to define samples of initial conditions of orbits whose properties will be identified. Fore this purpose, we define for several values of the total orbital energy $E$, dense uniform grids of $1024 \times 1024$ initial conditions regularly distributed on the $(x,y)$ plane inside the area allowed by the value of the energy. Following a typical approach, the orbits are launched with initial conditions inside a certain region, called scattering region, which in our case is a square grid with $-2\leq x,y \leq 2$.

\begin{figure*}[!tH]
\centering
\resizebox{0.6\hsize}{!}{\includegraphics{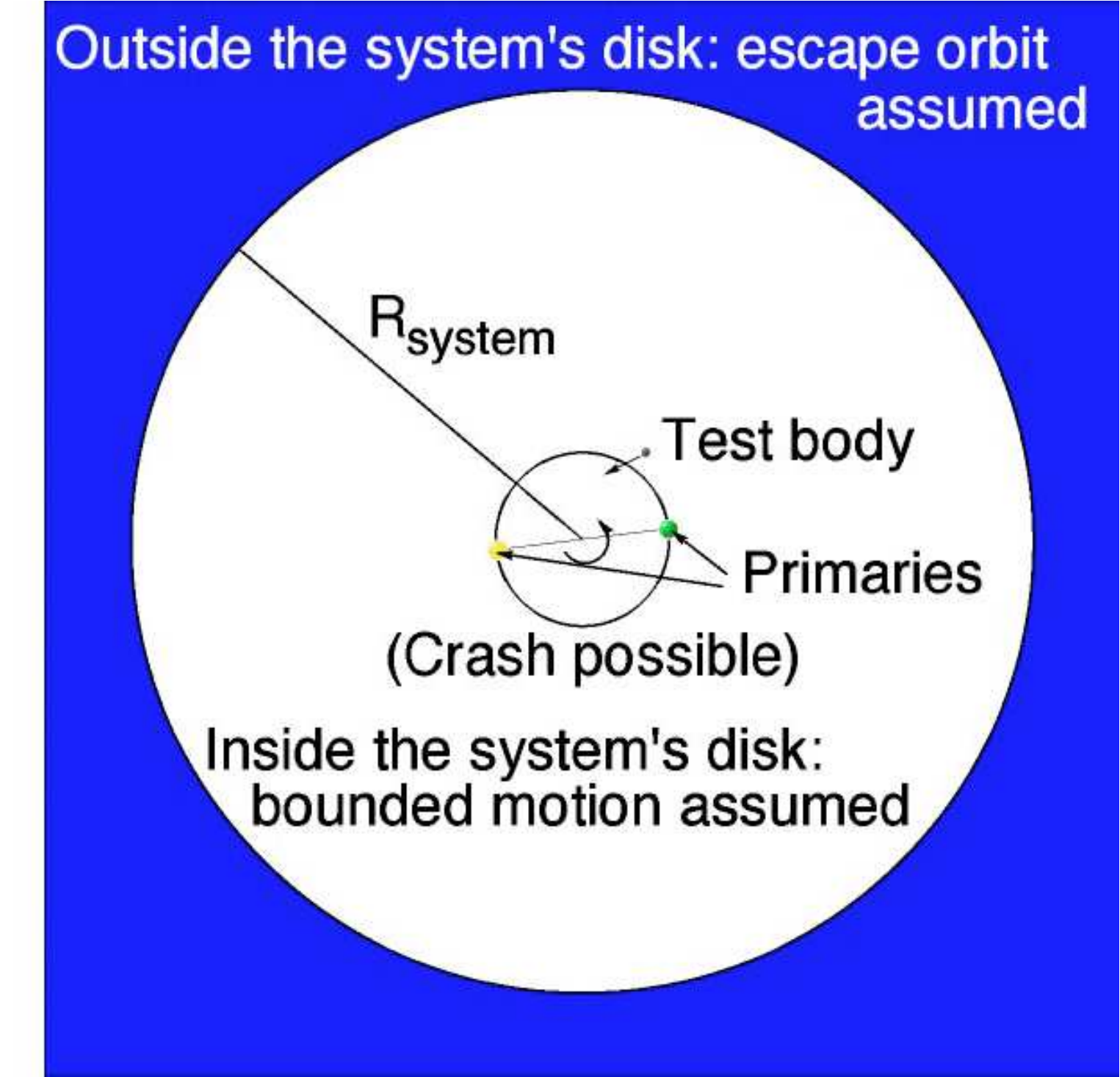}}
\caption{Schematic picture of the three different types of motion. The motion is considered to be bounded if the test body stays confined for integration time $t_{\rm max}$ inside the system's disk with radius $R_d = 10$, while the motion is unbounded and the numerical integration stops when the test body crosses the system's disk with velocity pointing outwards. Crash with one of the primaries occurs when the test body crosses the disk of radius $R_{m1} = R_{m2} = 10^{-4}$ of one of the primaries.}
\label{crit}
\end{figure*}

In the PCRTBP system the configuration space extends to infinity thus making the identification of the type of motion of the test body for specific initial conditions a rather demanding task. There are three possible types of motion for the test body: (i) bounded motion around one of the primaries, or even around both; (ii) escape to infinity; (iii) crash into one of the primaries. Now we need to define appropriate numerical criteria for distinguishing between these three types of motion. The motion is considered as bounded if the test body stays confined for integration time $t_{\rm max}$ inside the system's disk with radius $R_d$ and center coinciding with the center of mass origin at $(0,0)$. Obviously, the higher the values of $t_{\rm max}$ and $R_d$ the more plausible becomes the definition of bounded motion and in the limit $t_{\rm max} \rightarrow \infty$ the definition is the precise description of bounded motion in a finite disk of radius $R_d$. Consequently, the higher these two values, the longer the numerical integration of initial conditions of orbits lasts. In our calculations we choose $t_{\rm max} = 10^4$ and $R_d = 10$ (see Fig. \ref{crit}). We decided to include a relatively high disk radius $(R_d = 10)$ in order to be sure that the orbits will certainly escape from the system and not return back to the interior region. Furthermore, it should be emphasized that for low values of $t_{\rm max}$ the fractal boundaries of stability islands corresponding to bounded motion become more smooth. Moreover, an orbit is identified as escaping and the numerical integration stops if the test body body intersects the system's disk with velocity pointing outwards at a time $t_{\rm esc} < t_{\rm max}$. Finally, a crash with one of the primaries occurs if the test body, assuming it is a point mass, crosses the disk with radius $R_m$ around the primary, where in our case we choose $R_m = 10^{-4}$. Here is should be noted that in generally it is assumed that the radius of a celestial body (e.g., a planet) is directly proportional to the cubic root of its mass. For the sake of simplicity of the numerical calculations we decided to fix the radii of the primaries to $10^{-4}$ (see also \citet{BBS06}). In papers I and II it was shown that the radii of the primaries influence the area of crash and escape basins.

\begin{figure*}[!tH]
\centering
\resizebox{0.8\hsize}{!}{\includegraphics{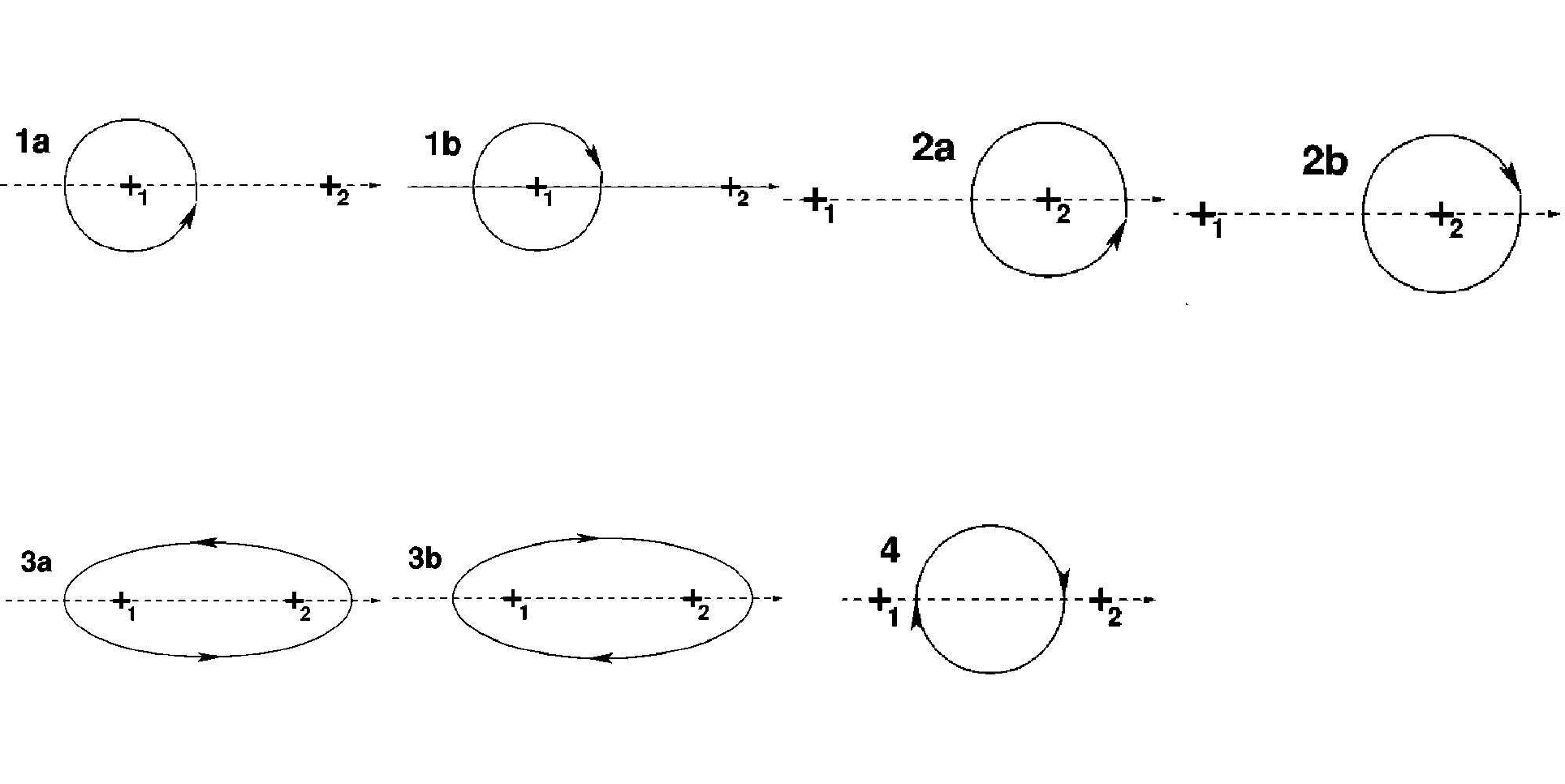}}
\caption{Characteristic orbit examples of the seven main types of regular orbits.}
\label{types}
\end{figure*}

The vast majority of bounded motion corresponds to initial conditions of regular orbits. It therefore seems appropriate to further classify initial conditions of ordered orbits into regular families. For this task we use the symbolic orbit classification which was also used in Papers I and II. According to this method orbits are classified by taking into account their orientation with respect to the centers of the two primaries $C_1$ and $C_2$, as well as their rotation (clockwise or counterclockwise). In particular, the orbit classification is based on an automatic detection of $x$ axis passages of the test body. Furthermore, two consecutive $x$ axis passages define a half rotation with respect to the fixed centers of the two primary bodies. In Fig. \ref{types} we present characteristic orbit examples of the seven main types of regular orbits. A more precise description of the types of orbits can be found in \citet{N02}.

As it was stated earlier, in our computations, we set $10^4$ time units as a maximum time of numerical integration. The vast majority of escaping orbits (regular and chaotic) however, need considerable less time to escape from the system (obviously, the numerical integration is effectively ended when an orbit moves outside the system's disk and escapes). Nevertheless, we decided to use such a vast integration time just to be sure that all orbits have enough time in order to escape. Remember, that there are the so called ``sticky orbits" which behave as regular ones during long periods of time. Here we should clarify, that orbits which do not escape after a numerical integration of $10^4$ time units are considered as non-escaping or trapped.

The equations of motion (\ref{eqmot}) for the initial conditions of all orbits are forwarded integrated using a double precision Bulirsch-Stoer \verb!FORTRAN 77! algorithm (e.g., \citet{PTVF92}) with a small time step of order of $10^{-2}$, which is sufficient enough for the desired accuracy of our computations. Here we should emphasize, that our previous numerical experience suggests that the Bulirsch-Stoer integrator is both faster and more accurate than a double precision Runge-Kutta-Fehlberg algorithm of order 7 with Cash-Karp coefficients. Throughout all our computations, the Jacobian energy integral (Eq. (\ref{ham})) was conserved better than one part in $10^{-11}$, although for most orbits it was better than one part in $10^{-12}$. For collisional orbits where the test body moves inside a region of radius $10^{-2}$ around one of the primaries the Lemaitre's global regularization method is applied.

\section{Numerical results \& Orbit classification}
\label{numres}

The main objective of our investigation is to classify initial condition of orbits in the physical $(x,y)$ plane into three categories: (i) bounded orbits; (ii) escaping orbits and (iii) crashing orbits, distinguishing simultaneously regular orbits into different types. Furthermore, two additional properties of the orbits will be examined: (i) the time-scale of crash and (ii) the time-scale of the escapes (we shall also use the terms escape period or escape rates). In the present paper, we shall explore these dynamical quantities for various values of the total orbital energy, as well as for the oblateness coefficient $A_1$. In particular, three different cases are considered: (a) the primary body with center at $C_1$ has a small value of $A_1$, (b) the primary body has an intermediate value of $A_1$ and (c) the primary body is highly oblate. In the following color-coded grids (or orbit type diagrams - OTDs) each pixel is assigned a color according to the orbit type. Thus the initial conditions of orbits are classified into bounded motion of a few types, unbounded escaping motion and collisional motion. In this special type of Poincar\'{e} surface of section the phase space emerges as a close and compact mix of escape basins, crash basins and stability islands.

Our numerical calculations indicate that apart from the escaping and crashing orbits there is also a considerable amount of non-escaping orbits. In general terms, the majority of non-escaping regions corresponds to initial conditions of regular orbits, where an adelphic integral of motion is present, restricting their accessible phase space and therefore hinders their escape. Additional numerical computations reveal that the bounded regular orbits are mainly loop 1:1 resonant orbits for which the adelphic integral applies\footnote{The total angular momentum is an approximately conserved quantity (integral of motion), even for orbits in non spherical potentials.}, while other types of secondary resonant orbits are also present. The $n:m$ notation we use for the regular orbits is according to \citet{CA98} and \citet{ZC13}, where the ratio of those integers corresponds to the ratio of the main frequencies of the orbit, where main frequency is the frequency of greatest amplitude in each coordinate. Main amplitudes, when having a rational ratio, define the resonances of an orbit.

\subsection{Case I: Results for a low value of oblateness}
\label{cas1}

\begin{figure*}[!tH]
\centering
\resizebox{0.85\hsize}{!}{\includegraphics{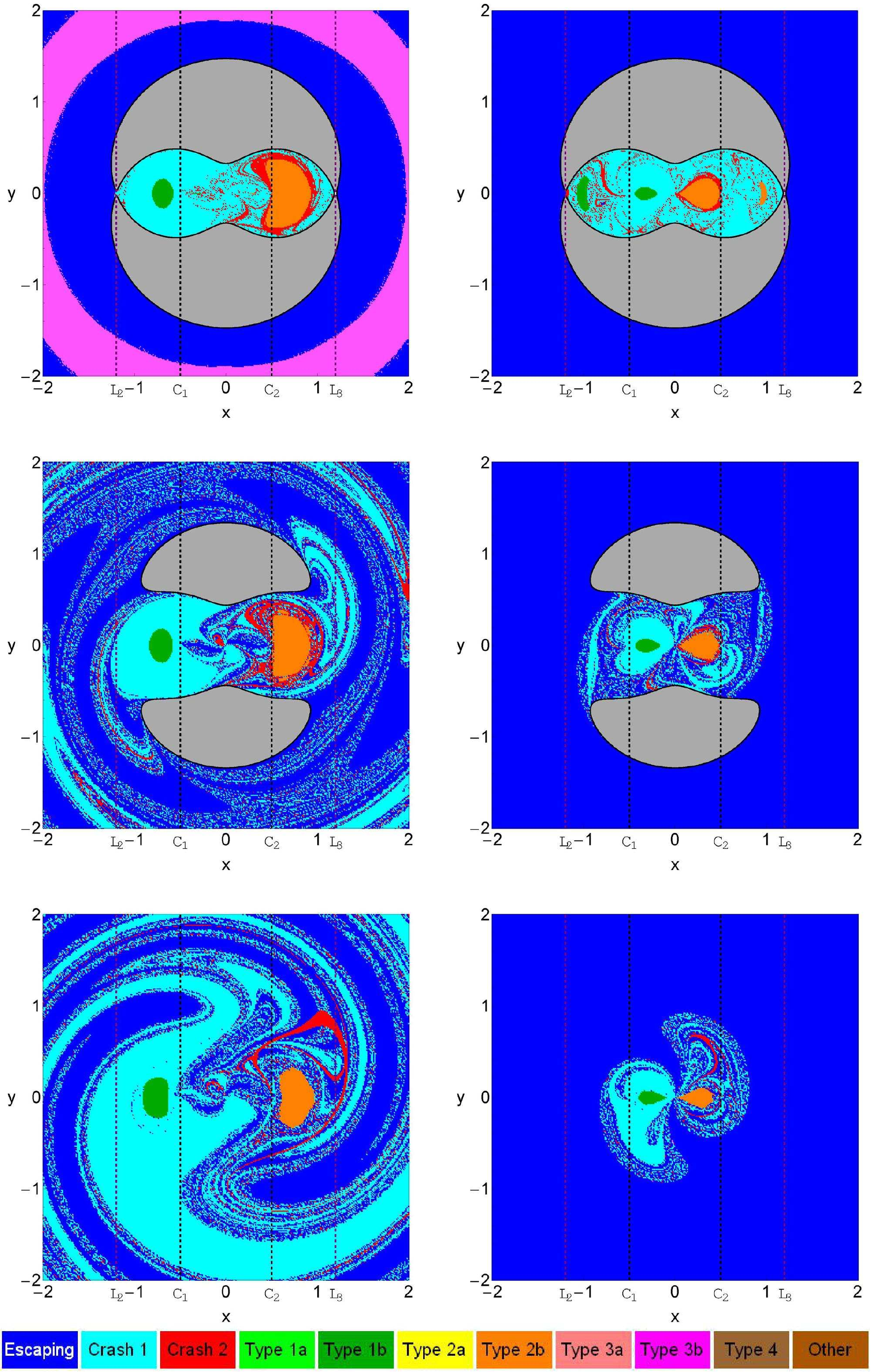}}
\caption{The orbital structure of the physical $(x,y)$ plane in a corotating frame of reference is given using orbit type diagrams (OTDs) for three energy levels and for both parts $\dot{\phi} < 0$ (left column) and $\dot{\phi} > 0$ (right column) of the surface of section $\dot{r} = 0$, when $A_1 = 0.001$. (Top row): $E = -1.73$; (middle row): $E = -1.60$; (bottom row): $E = -1.30$. The vertical black dashed lines denote the centers of the two primaries, wile the vertical purple dashed lines indicate the position of the Lagrangian points $L_2$ and $L_3$. The color bar contains the color code which relates the types of orbits presented in Fig. \ref{types} with different colors.}
\label{grd1}
\end{figure*}

\begin{figure*}
\centering
\resizebox{\hsize}{!}{\includegraphics{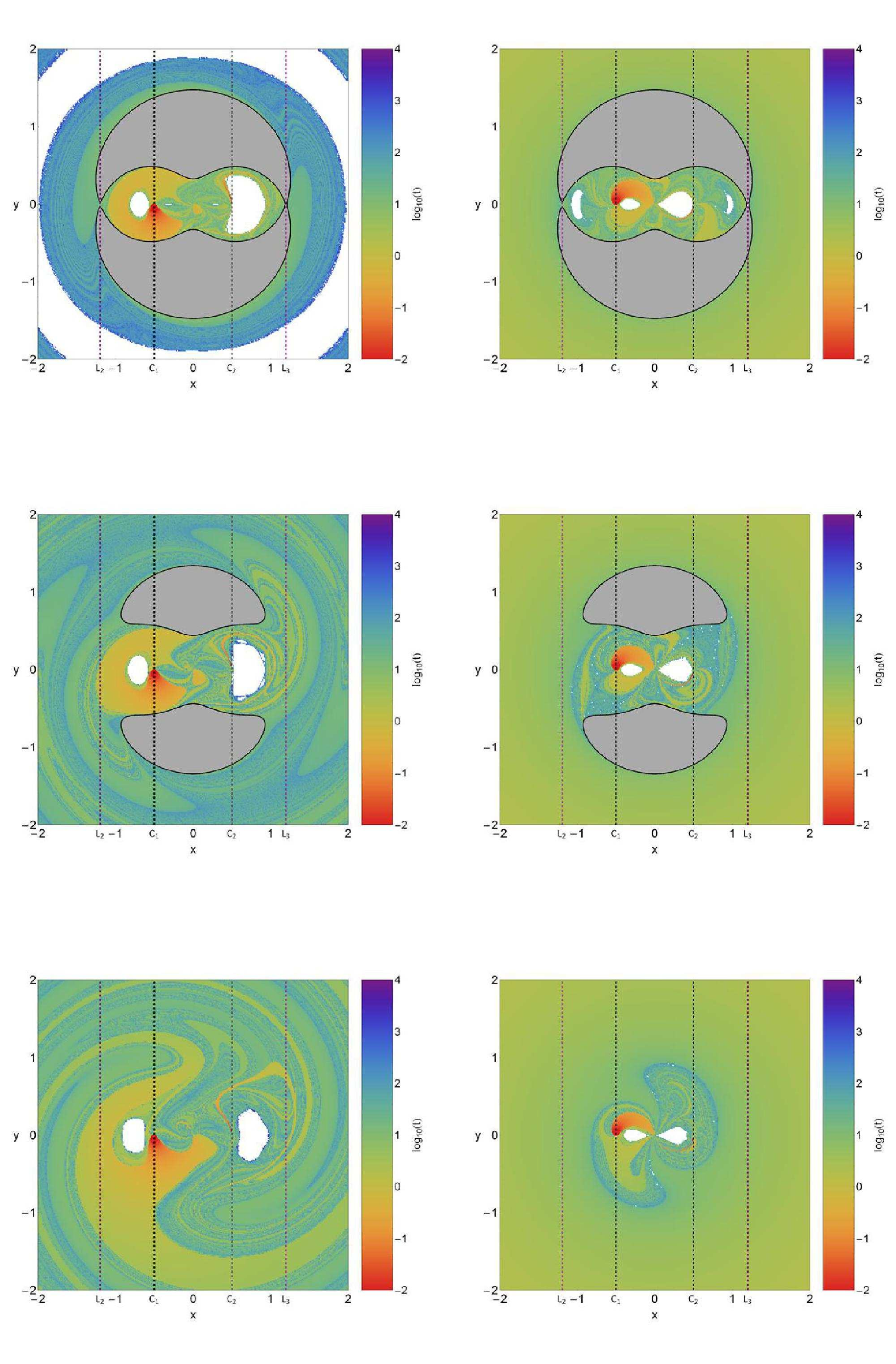}}
\caption{Distribution of the escape and collisional time of the orbits on the physical $(x,y)$ space when $A_1 = 0.001$ for the energy levels of Fig. \ref{grd1}. (Top row): $E = -1.73$; (middle row): $E = -1.60$; (bottom row): $E = -1.30$. The darker the color, the larger the escape/crash time. Initial conditions of bounded regular orbits are shown in white.}
\label{t1}
\end{figure*}

\begin{figure*}
\centering
\resizebox{\hsize}{!}{\includegraphics{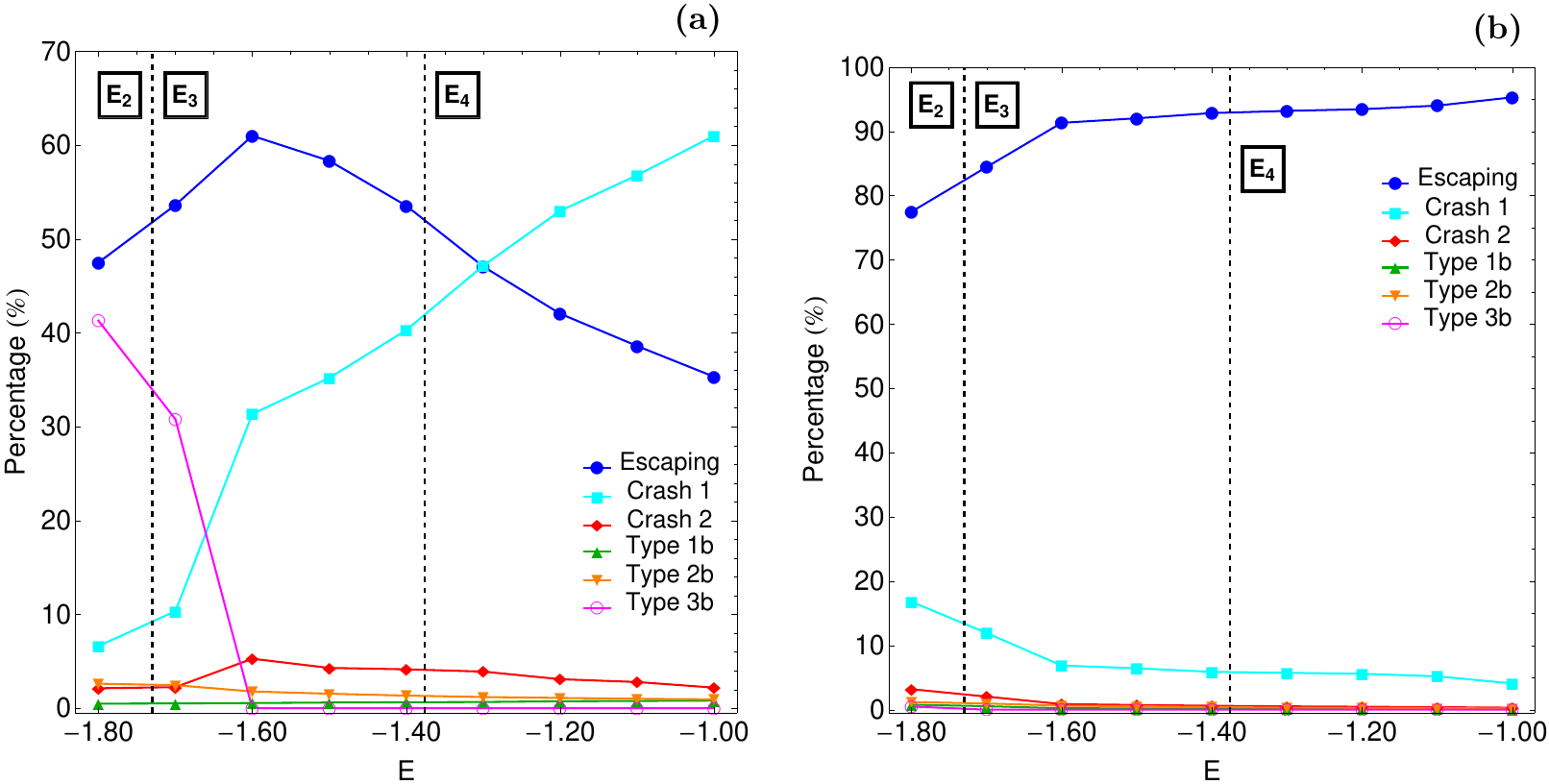}}
\caption{Evolution of the percentages of all types of orbits on the physical $(x,y)$ plane, for $A_1 = 0.001$ when varying the energy parameter $E$ for (a-left): the $\dot{\phi} < 0$ part and (b-right): the $\dot{\phi} > 0$ part of the configuration space.}
\label{p1}
\end{figure*}

Our exploration begins considering the case where the primary body 1 has a relatively low value of oblateness, that is the case of $A_1 = 0.001$. In Fig. \ref{grd1} the OTD decompositions for both $\dot{\phi} < 0$ (left column) and $\dot{\phi} > 0$ (right column) reveal the structure of the physical $(x,y)$ space for three energy levels, where the several types of orbits are indicated with different colors. The color code is explained in the color bar at the bottom of the figure. The black solid lines in the two types of plots denote the Zero Velocity Curve, while the inaccessible forbidden regions are marked in gray. The color of a point represents the orbit type of a test body which has been launched with pericenter position at $(x,y)$. The three energy levels belong to each of the three Hill's regions configurations explained earlier in Fig. \ref{conts}. The cases $E < E_2$ and $E_2 < E < E_3$ give very similar results, so we included only the latter. When $E = -1.73$ we observe that in both cases the interior region is filled with initial conditions of orbits that either are regular or crash into one of the primaries, while initial conditions of escaping orbits are present only in the exterior region outside $L_2$ and $L_3$. Regular motion dominates the interior region and two large stability islands are shown near the centers of the primaries. These stability islands contain quasi-periodic orbits which are symmetrical with respect to a reflection over the $x$ axis and they move in clockwise sense, hence, retrograde in relation to the rotating system of coordinates. Moreover, the stability regions are surrounded by a mix of domains of crash orbits with respect to the first and the second primary body. On the other hand, the exterior region contains mostly initial conditions of escaping orbits however, in the $\dot{\phi} < 0$ plot we have to point out the existence of a stability ring containing initial conditions of regular orbits that circulate clockwise around both primaries. As the value of the energy increases and the area of forbidden region is reduced it is seen that the stability ring disappears, while in both the interior and exterior regions basins of escaping and collisional orbits are formed. We should note though, that the basins containing orbits that crash into primary 2 are much smaller, with respect to the crash basins of primary 1, and they appear only as thin filaments. This phenomenon is justified by the high value of the oblateness of primary 1. For $E = -1.30 > E_4$ the test body has full access to the entire physical $(x,y)$ plane. Two regions of bounded motion are shown around the primaries, where for each stability island there is a parent periodic orbit at its center. Furthermore, due to the rotation of the primaries the crash basins wind out in spiral form in the outer regions of the $\dot{\phi} < 0$ plot. Crash basins are also present in the immediate vicinity of the origin. At this point we would like to stress out that the area of crash basins is several orders of magnitude larger than the total size of the primary body's disks. In the OTD for $\dot{\phi} > 0$ the escape basin cover the vast majority of the configuration space, while the total size of stability regions is less than for $\dot{\phi} < 0$. It is interesting to note that all regular orbits in all three energy levels were found to be retrograde thus traveling in clockwise sense.

The following Fig. \ref{t1} shows how the escape and crash times of orbits are distributed on the physical $(x,y)$ space for the three energy levels discussed in Fig. \ref{grd1}. Light reddish colors correspond to fast escaping/crashing orbits, dark blue/purple colors indicate large escape/crash rates, while white color denote stability islands of regular motion. Note that the scale on the color bar is logarithmic. Inspecting the spatial distribution of various different ranges of escape time, we are able to associate medium escape time with the stable manifold of a non-attracting chaotic invariant set, which is spread out throughout this region of the chaotic sea, while the largest escape time values on the other hand, are linked with sticky motion around the stability islands of the two primaries. It should be noted that the behaviour of the escape times is very similar to that observed in \citet{dAT14}. As for the collision time we see that orbits with initial conditions very close to the vicinity of the center of the oblate primary 1 collide with it almost immediately, within the first time step of the numerical integration.

The evolution of the percentages of the different types of orbits on the physical $(x,y)$ space for both parts $\dot{\phi} < 0$ (a) and $\dot{\phi} > 0$ (b) when the energy varies is presented in Fig. \ref{p1}(a-b). The vertical black dashed lines indicate the three critical values of the Jacobi integral $(E_2, E_3, E_4)$. It is seen that for $A_1 = 0.001$, $E_2$ and $E_3$ are very close and therefore the corresponding lines are almost indistinguishable. One may observe that in both cases three types of orbits control the vast majority of the configuration space: (i) escaping orbits; (ii) orbits which crash into oblate primary 1 and (iii) type 3b regular orbits that is orbits which circulate clockwise around both primaries, while all the rest types are almost unaffected by the energy shifting, having considerably low rates (less than 5\%) throughout. For the $\dot{\phi} < 0$ case we see that for low values of energy $E < E_2$, escaping and type 3b regular orbits seem to share about 90\% of the physical space, while initial conditions of orbits that crash into oblate primary 1 occupy only about 7\% of the same plane. As the value of the energy increases however, the percentage of orbits that crash into the oblate primary increases and for $E > -1.3$ it dominates, while the rate of type 3b regular orbits on the other hand, reduces drastically and for $E > -1.6$ it completely vanishes. Moreover, the percentage of escaping orbits increases linearly for $E < -1.6$, while for larger values of the energy this tendency is reversed. At the highest energy level studied $(E = -1)$, about 35\% of the physical space is occupied by escaping orbits, while initial conditions of orbits that lead to collision with primary 1 correspond to about 60\% of the $(x,y)$ plane. For the the $\dot{\phi} > 0$ case, things are much more simpler since always the vast majority (more than 80\%) of the configuration space is dominated by escaping orbits, while the rate of collisional orbits to primary 1 is significantly lower; it starts at about 20\% at low values of energy $(E < E_2)$ and drops to about 5\% for $E > E_4$.

\subsection{Case II: Results for an intermediate value of oblateness}
\label{cas2}

\begin{figure*}[!tH]
\centering
\resizebox{0.85\hsize}{!}{\includegraphics{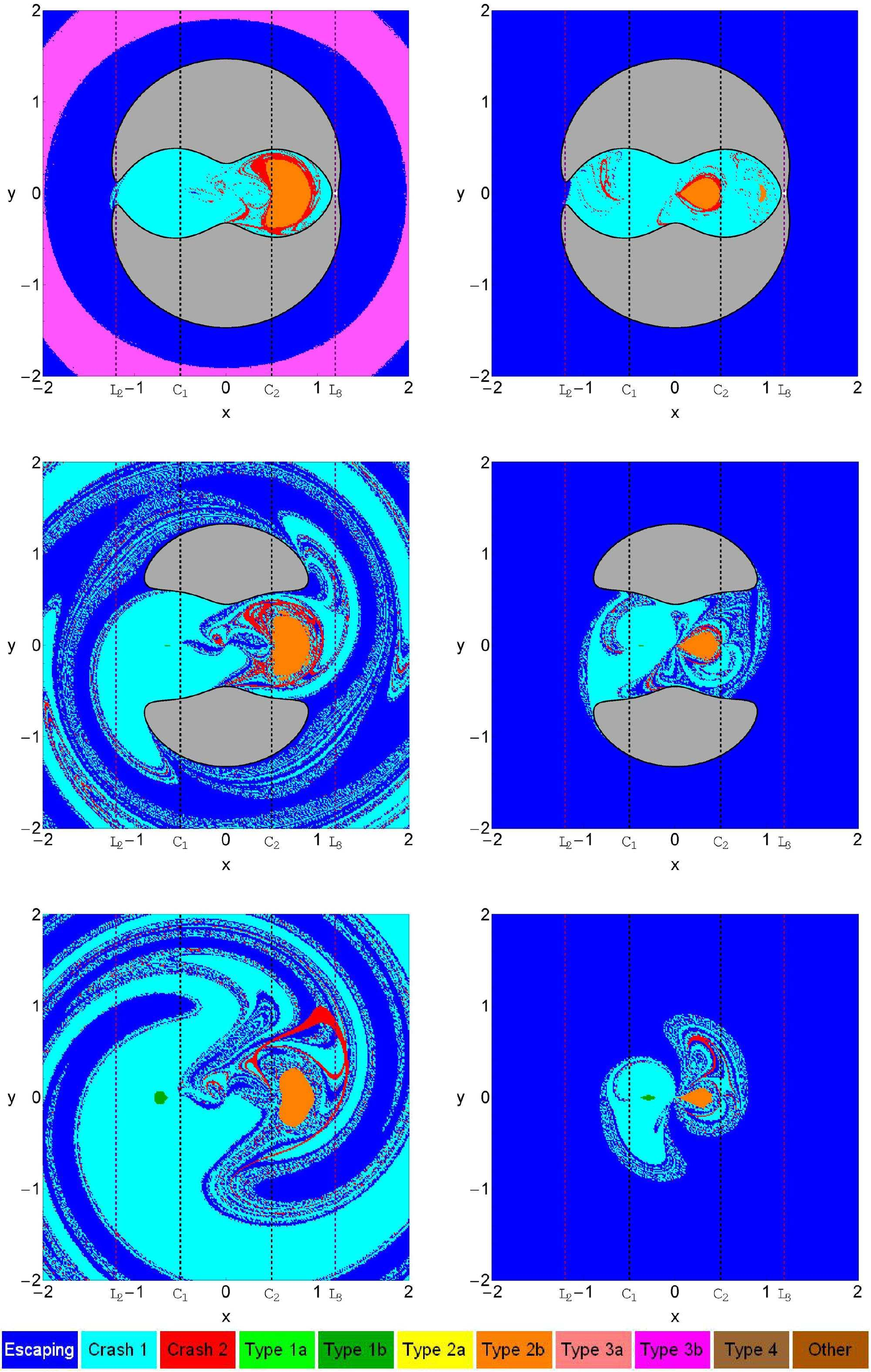}}
\caption{The orbital structure of the physical $(x,y)$ plane in a corotating frame of reference is given using orbit type diagrams (OTDs) for three energy levels and for both parts $\dot{\phi} < 0$ (left column) and $\dot{\phi} > 0$ (right column) of the surface of section $\dot{r} = 0$, when $A_1 = 0.01$. (Top row): $E = -1.742$; (middle row): $E = -1.60$; (bottom row): $E = -1.30$. The vertical black dashed lines denote the centers of the two primaries, wile the vertical purple dashed lines indicate the position of the Lagrangian points $L_2$ and $L_3$. The color bar contains the color code which relates the types of orbits presented in Fig. \ref{types} with different colors.}
\label{grd2}
\end{figure*}

\begin{figure*}
\centering
\resizebox{\hsize}{!}{\includegraphics{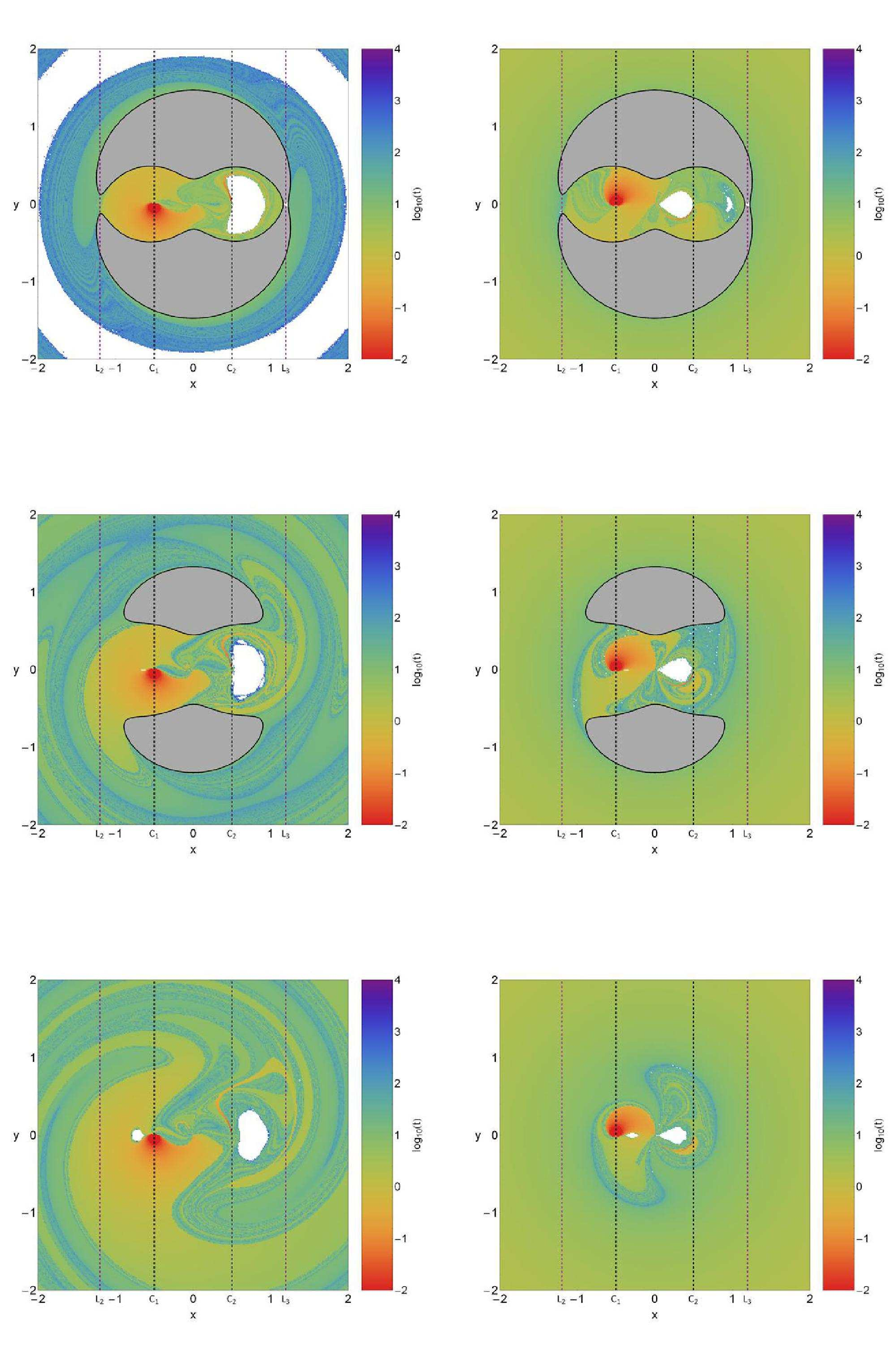}}
\caption{Distribution of the escape and collisional time of the orbits on the physical $(x,y)$ space when $A_1 = 0.01$ for the energy levels of Fig. \ref{grd2}. (Top row): $E = -1.742$; (middle row): $E = -1.60$; (bottom row): $E = -1.30$. The darker the color, the larger the escape/crash time. Initial conditions of bounded regular orbits are shown in white.}
\label{t2}
\end{figure*}

\begin{figure*}
\centering
\resizebox{\hsize}{!}{\includegraphics{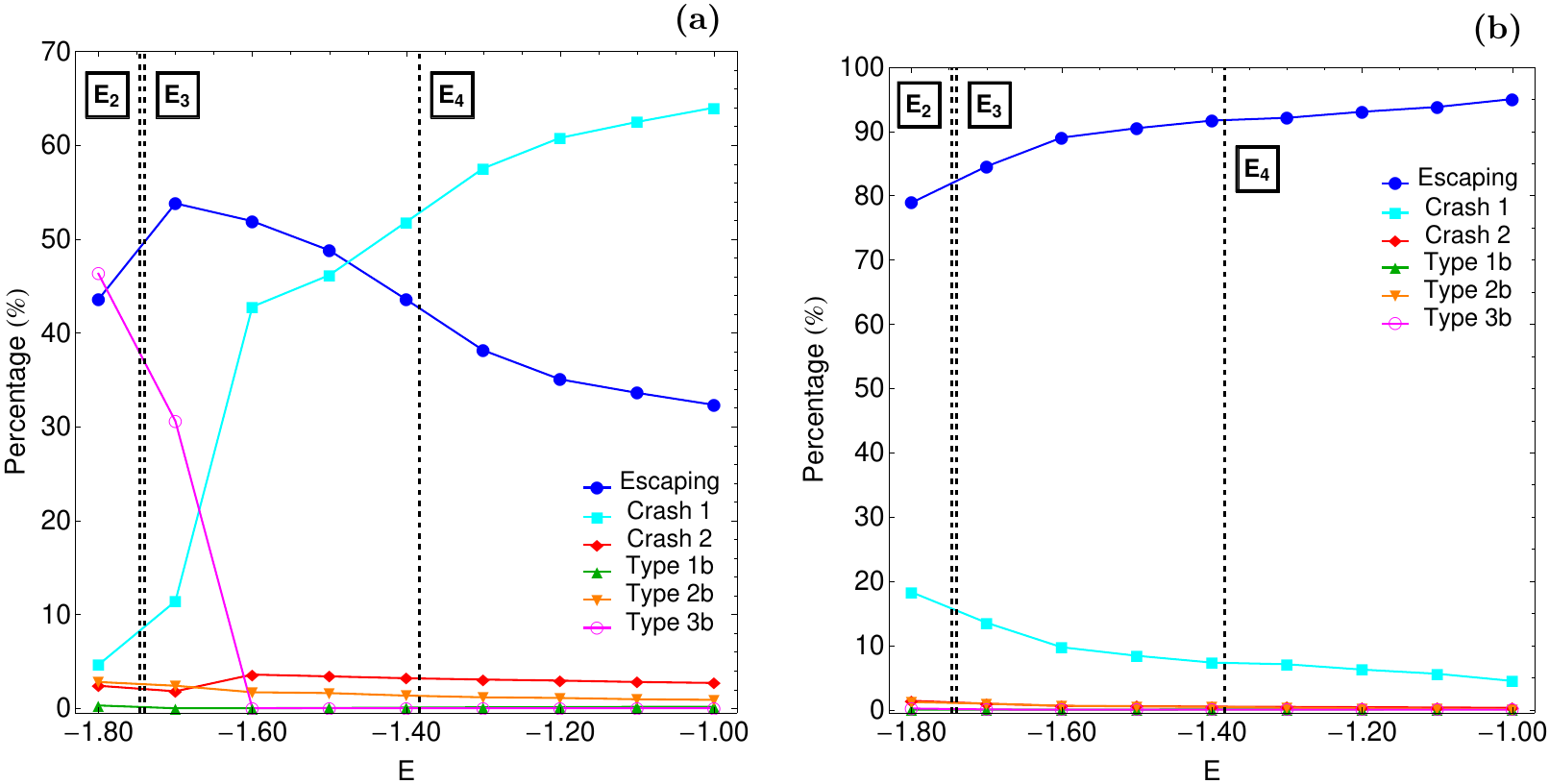}}
\caption{Evolution of the percentages of all types of orbits on the physical $(x,y)$ plane, for $A_1 = 0.01$ when varying the energy parameter $E$ for (a-left): the $\dot{\phi} < 0$ part and (b-right): the $\dot{\phi} > 0$ part of the configuration space.}
\label{p2}
\end{figure*}

We continue our numerical quest considering the case where the oblateness coefficient of the primary body 1 has an intermediate value, that is $A_1 = 0.01$, and we follow the same numerical approach as previously. The structure of the physical $(x,y)$ space for three energy levels is unveiled in Fig. \ref{grd2} through the OTD decompositions for both $\dot{\phi} < 0$ (left column) and $\dot{\phi} > 0$ (right column) parts of the configuration space. We observe that in general terms for both cases ($\dot{\phi} < 0$ and $\dot{\phi} > 0$), things are quite similar to those discussed earlier in Fig. \ref{grd1}. However there is one major difference regarding the stability regions. In particular, for the energy level $E = -1.742$, that is for the Hill's region $E_2 < E < E_3$, the stability island near the center of the oblate primary 1 is absent. At $E = -1.6$ a tiny stability region emerges at the left and right part of the center for $\dot{\phi} < 0$ and $\dot{\phi} > 0$, respectively, while only at a relatively high energy level $E = -1.3 > E_4$ bounded retrograde motion around primary 1 is possible. Therefore, we may conclude that the oblateness coefficient influences significantly the regular orbits around the oblate primary 1. Our numerical computations strongly indicate that the more oblate is primary 1 the less bounded motion is observed around it, while at the same time most of the initial conditions of orbits launched relatively close to primary 1 lead to fast collision with it. Moreover, as we seen in Fig. \ref{grd1} the $\dot{\phi} > 0$ part of the configuration space is dominated by escaping orbits, while inside the interior region crash with the left oblate primary body becomes more and more likely as we proceed to higher energy levels. Around the spherical (non-oblate) primary body 2 on the other hand, the area occupied by initial conditions corresponding to bounded motion around primary 2 exhibits a minor decrease with increasing energy. In addition, the crash basin to primary 2 is much smaller with respect to the extended crash basin 1, and has a spiral shape around primary 2. The distribution of the escape and crash times of orbits on both parts of the physical space is shown in Fig. \ref{t2}. One may observe that the results are very similar to those presented earlier in Fig. \ref{t1}, where we found that orbits with initial conditions inside the escape and crash basins have the smallest escape/crash rates, while on the other hand, the longest escape/crash times correspond to orbits with initial conditions in the fractal regions of the plots. It is interesting to note that orbits with initial conditions in the vicinity of the center of the oblate primary 1 collide with it almost immediately, while this phenomenon is not observed for the case of primary body 2 which is not oblate.

In Fig. \ref{p2}(a-b) we demonstrate the evolution of the percentages of the different types of orbits on the physical $(x,y)$ space for both parts $\dot{\phi} < 0$ (a) and $\dot{\phi} > 0$ (b) as a function of the total orbital energy. We observe that for $A_1 = 0.01$ the vertical black dashed lines indicating the critical values of the energy $E_2$ and $E_3$ are no longer indistinguishable as it was in Fig. \ref{p1} for $A_1 = 0.001$. Once more it is seen that in both parts of the configuration space only three types of orbits (escaping, crash into primary 1 and regular type 3b) are influenced by the change on the value of the energy, while all the other types of orbits seem completely unaffected by the energy shifting holding very low rates (less than about 5\% throughout the energy range). In Fig. \ref{p2}a we see that for low value of the energy $E < E_2$ the percentages of escaping and type 3b regular orbits seem to coincide at about 45\%, while for larger energies they follow completely different paths. In particular, the rate of escaping orbits initially increases but for $E > E_3$ it exhibits a constant and almost linear decrease, while the rate of type 3b ordered orbits displays a rapid reduction and for $E > -1.6$ it completely vanishes. The percentage of orbits that crash into primary 1 on the other hand, increases drastically and for $E > -1.5$ this type of orbits is the most populated family. Moreover, at the highest energy level studied, that is $E = -1$, crashing into primary 1 orbits cover about two thirds of the entire physical $(x,y)$ space. The orbital structure of the $\dot{\phi} > 0$ part of the configuration space is completely different. Indeed, in Fig. \ref{p2}b one may observe that escaping orbits dominate the physical plane throughout the energy range. For low values of the energy the corresponding rate ia about 80\%, while for $E > E_4$ it exceeds 95\%. The evolution of the percentage of orbits that crash into oblate primary 1 follows an opposite pattern, starting from about 20\% and drops to about 5\% for $E > -1.2$. We should point out that in this case, the rates of all the other types of orbits are always less than 2\%, while the change on the value of the energy only shuffles the orbital content among them.

\subsection{Case III: Results for a high value of oblateness}
\label{cas3}

\begin{figure*}[!tH]
\centering
\resizebox{0.85\hsize}{!}{\includegraphics{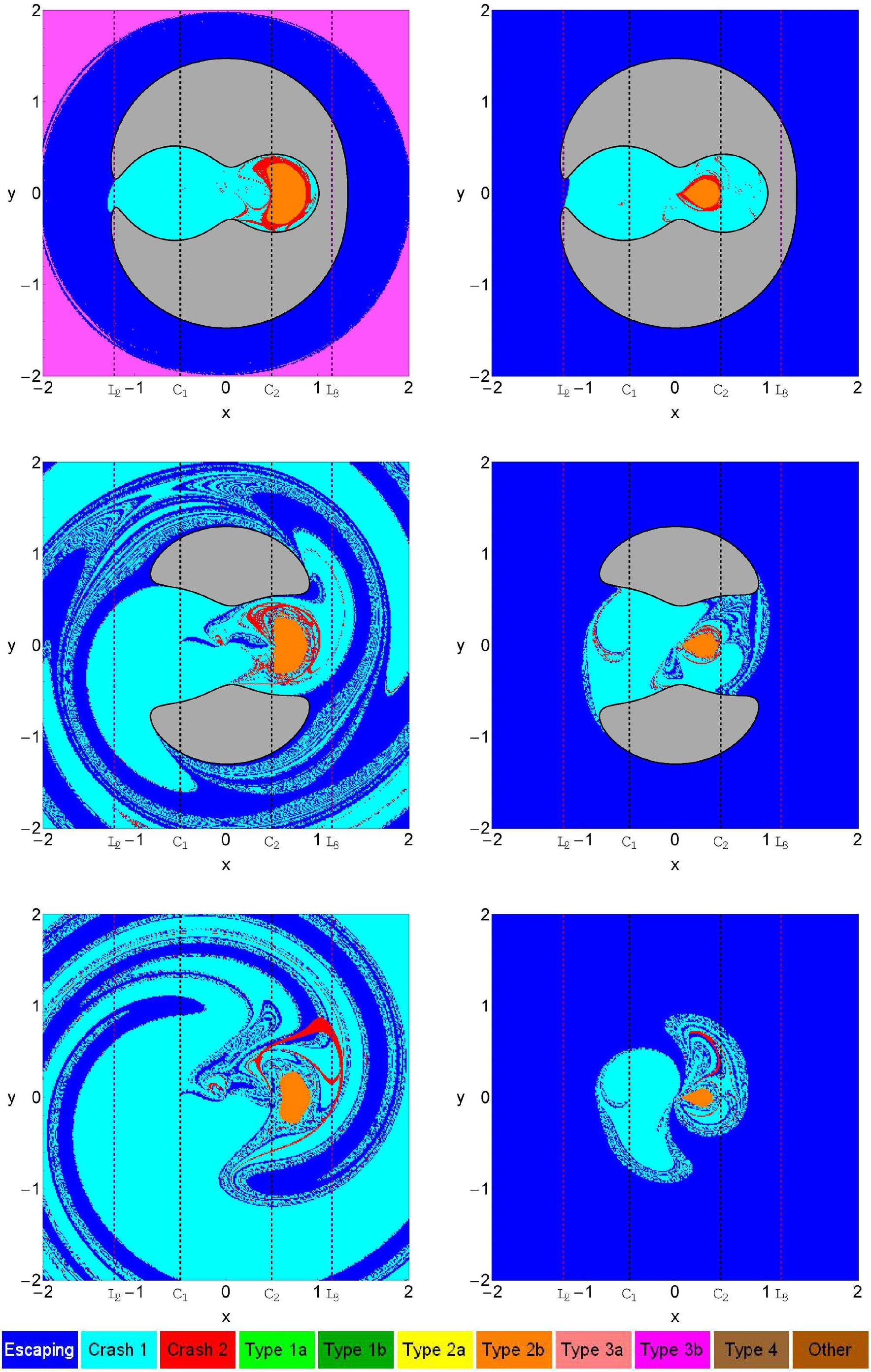}}
\caption{The orbital structure of the physical $(x,y)$ plane in a corotating frame of reference is given using orbit type diagrams (OTDs) for three energy levels and for both parts $\dot{\phi} < 0$ (left column) and $\dot{\phi} > 0$ (right column) of the surface of section $\dot{r} = 0$, when $A_1 = 0.1$. (Top row): $E = -1.90$; (middle row): $E = -1.70$; (bottom row): $E = -1.40$. The vertical black dashed lines denote the centers of the two primaries, wile the vertical purple dashed lines indicate the position of the Lagrangian points $L_2$ and $L_3$. The color bar contains the color code which relates the types of orbits presented in Fig. \ref{types} with different colors.}
\label{grd3}
\end{figure*}

\begin{figure*}
\centering
\resizebox{\hsize}{!}{\includegraphics{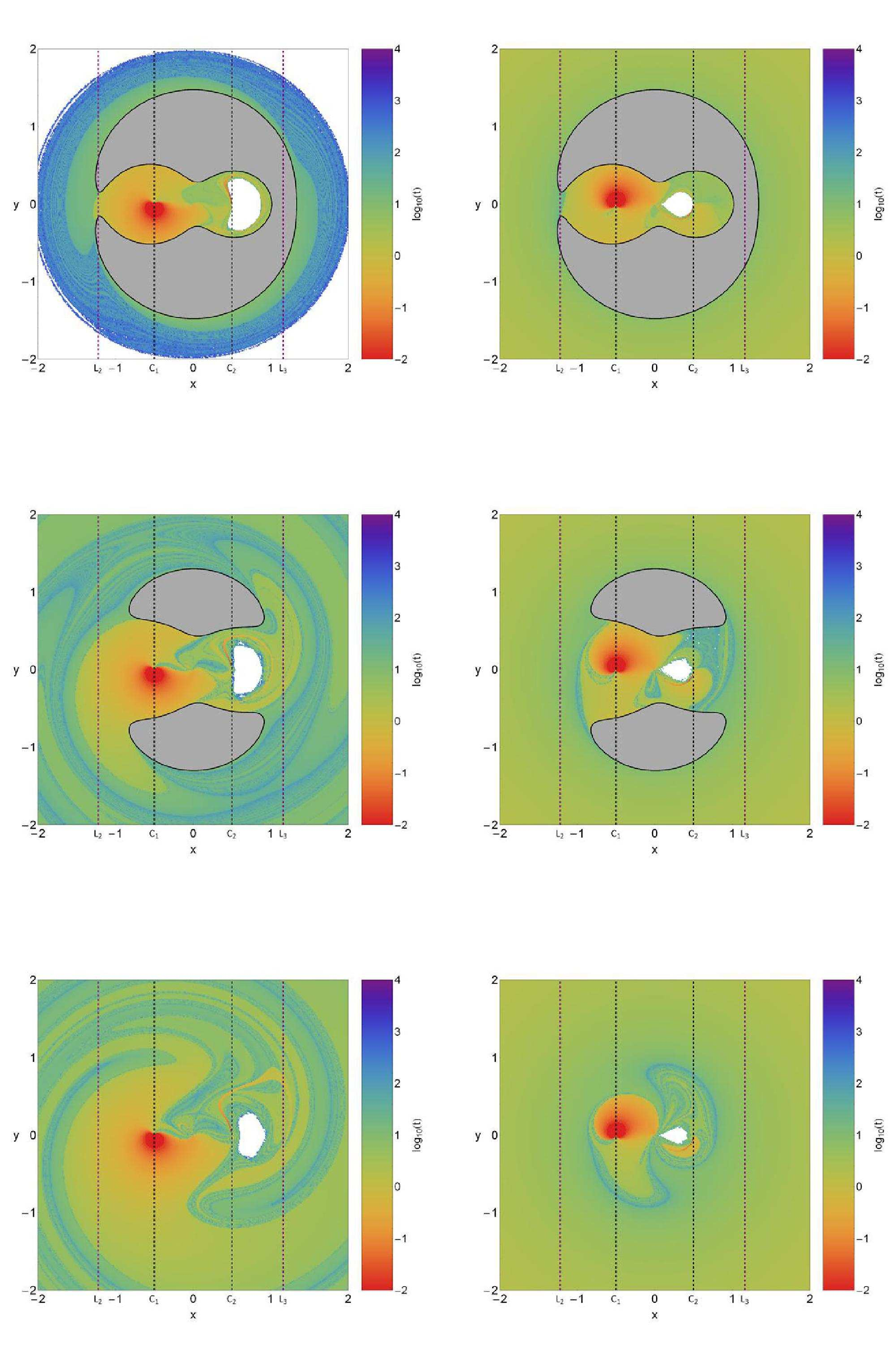}}
\caption{Distribution of the escape and collisional time of the orbits on the physical $(x,y)$ space when $A_1 = 0.1$ for the energy levels of Fig. \ref{grd3}. (Top row): $E = -1.90$; (middle row): $E = -1.70$; (bottom row): $E = -1.40$. The darker the color, the larger the escape/crash time. Initial conditions of bounded regular orbits are shown in white.}
\label{t3}
\end{figure*}

\begin{figure*}
\centering
\resizebox{\hsize}{!}{\includegraphics{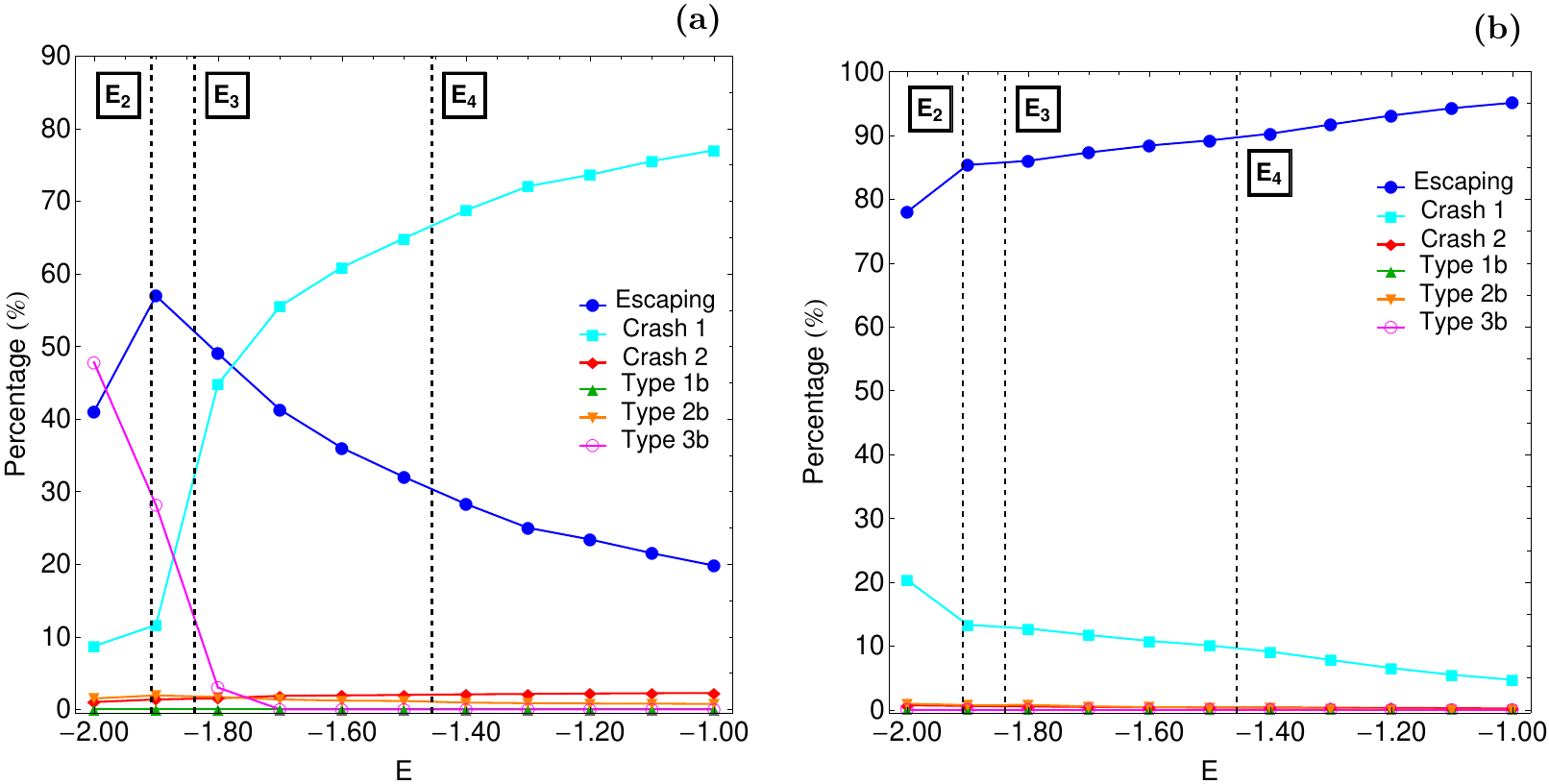}}
\caption{Evolution of the percentages of all types of orbits on the physical $(x,y)$ plane, for $A_1 = 0.1$ when varying the energy parameter $E$ for (a-left): the $\dot{\phi} < 0$ part and (b-right): the $\dot{\phi} > 0$ part of the configuration space.}
\label{p3}
\end{figure*}

The last case under investigation involves the scenario where the oblateness coefficient of the primary body 1 has a relatively high value, that is $A_1 = 0.1$. Again, all the different aspects of the numerical approach remain exactly the same as in the two previously studied cases. The orbital structure of the physical $(x,y)$ plane through the OTD decompositions for both $\dot{\phi} < 0$ (left column) and $\dot{\phi} > 0$ (right column) parts of the configuration space distinguishing between the main types of orbits for three energy levels is shown in Fig. \ref{grd3}. For the $\dot{\phi} < 0$ part it is evident that when $E = -1.90 < E_3$ all the interior region is occupied either by collisional orbits or regular orbits, while the exterior region is divided into two domains; a circular domain containing initial conditions of escaping orbits and another domain covered by type 3b ordered orbits that circulate clockwise around both primaries. The collisional to primary 2 basin is appeared as thin layer outside the main stability island. As the value of the energy increases the extent of this stability island is reduced, while a considerable amount of initial conditions of orbits located inside the interior region lead to escape. At the same time, a large portion of the exterior region is occupied by extended crash basins spiralling around the center of the coordinates. The $\dot{\phi} > 0$ part of the configuration space displays, once more,  the same structure as in the previous two cases, where we found that the vast majority of the integrated initial conditions correspond to orbits which escape from the system. In the previous case where the primary body 1 had an intermediate value of oblateness, we deduced that the more oblate is one primary the less bounded motion is observed around it. This conclusion becomes more clear and strong here where we see for a high value of the oblateness ($A_1 = 0.1$) there is no indication of regular motion around primary 1 for all tested energy levels and in both parts of the configuration space. Thus we may certainly conclude that the increase on the oblateness of the primary body has a detrimental effect on the stability islands near the same primary, leading most test bodies in collisional orbits with the oblate primary. In Fig. \ref{t3} we depict the distribution of the escape/crash times of orbits for both parts of the physical space, where one can see similar outcomes with that presented in the two previous subsections. At this point, we would like to point out that the basins of escape can be easily distinguished in Fig. \ref{t3}, being the regions with intermediate greenish colors indicating fast escaping orbits. Indeed, our numerical calculations suggest that orbits with initial conditions inside these basins need no more than 10 time units to escape from the system. Furthermore, the crash basins are shown with reddish colors where the corresponding crash time is less than one time unit.

Finally, Fig. \ref{p3}(a-b) shows how the percentages of all types of orbits on both parts of the physical space evolve when the total orbital energy varies in the interval $E \in [-2,-1]$. In this case, the four energy intervals defined by the three critical values of the energy are fully distinguishable and are the following: (i) $E < E_2$; (ii) $E_2 < E < E_3$; (iii) $E_3 < E < E_4$; (iv) $E > E_4$. Fig. \ref{p3}a shows that in the first energy interval escaping and type 3b regular orbits occupy about 40\% and 50\% of the physical plane, respectively, while the rest area corresponds to initial conditions of orbits that crash into primary 1. For $E > E_2$ however, the rate of type 3b regular orbits drops suddenly and for $E > E_3$ it vanishes. In the same vein, the percentage of escaping orbits is about 60\% for $E = E_2$, while for larger values of energy it decreases reaching about one third of its initial value $(20\%)$ for $E = -1$. The percentage of orbits that crash into the oblate primary on the other hand, grows drastically for $E > E_2$ and at high values of energy $(E > E_4)$ it seems to saturate to around 80\%. Once more we observe, that the variation on the value of the total orbital energy has practically no influence on the rates of all the other types of orbits which remain unperturbed and at extremely low values (less than 5\%). We seen in Fig. \ref{grd3} that the $\dot{\phi} > 0$ part of the configuration space is dominated by initial conditions of orbits that escape from the system. Fig. \ref{p3}b verifies this claim since the corresponding percentage remains high (more than 80\%) throughout. Moreover one can observe that in this case crash on the oblate primary body becomes less and less unlike with increasing energy due to the fact that the rate of this family decreases as we proceed to higher energy levels and for $E = -1$ they occupy only about 5\% of the physical space.

Before closing this section we would like to emphasize that the OTDs given in Figs. \ref{grd1}, \ref{grd2} and \ref{grd3} have both fractal and non-fractal (smooth) boundary regions which separate the escape basins and the collisional basins. Such fractal basin boundaries is a common phenomenon in leaking Hamiltonian systems (e.g., \citet{BGOB88,dML99,dMG02,STN02,ST03,TSPT04}). In the PCRTBP system the leakages are defined by both escape and crash conditions thus resulting in three exit modes. However, due to the high complexity of the basin boundaries, it is very difficult, or even impossible, to predict in these regions whether the test body (e.g., a satellite, asteroid, planet etc) collides with a primary body or escapes from the dynamical system.

\section{An overview analysis}
\label{over}

\begin{figure*}[!tH]
\centering
\resizebox{0.85\hsize}{!}{\includegraphics{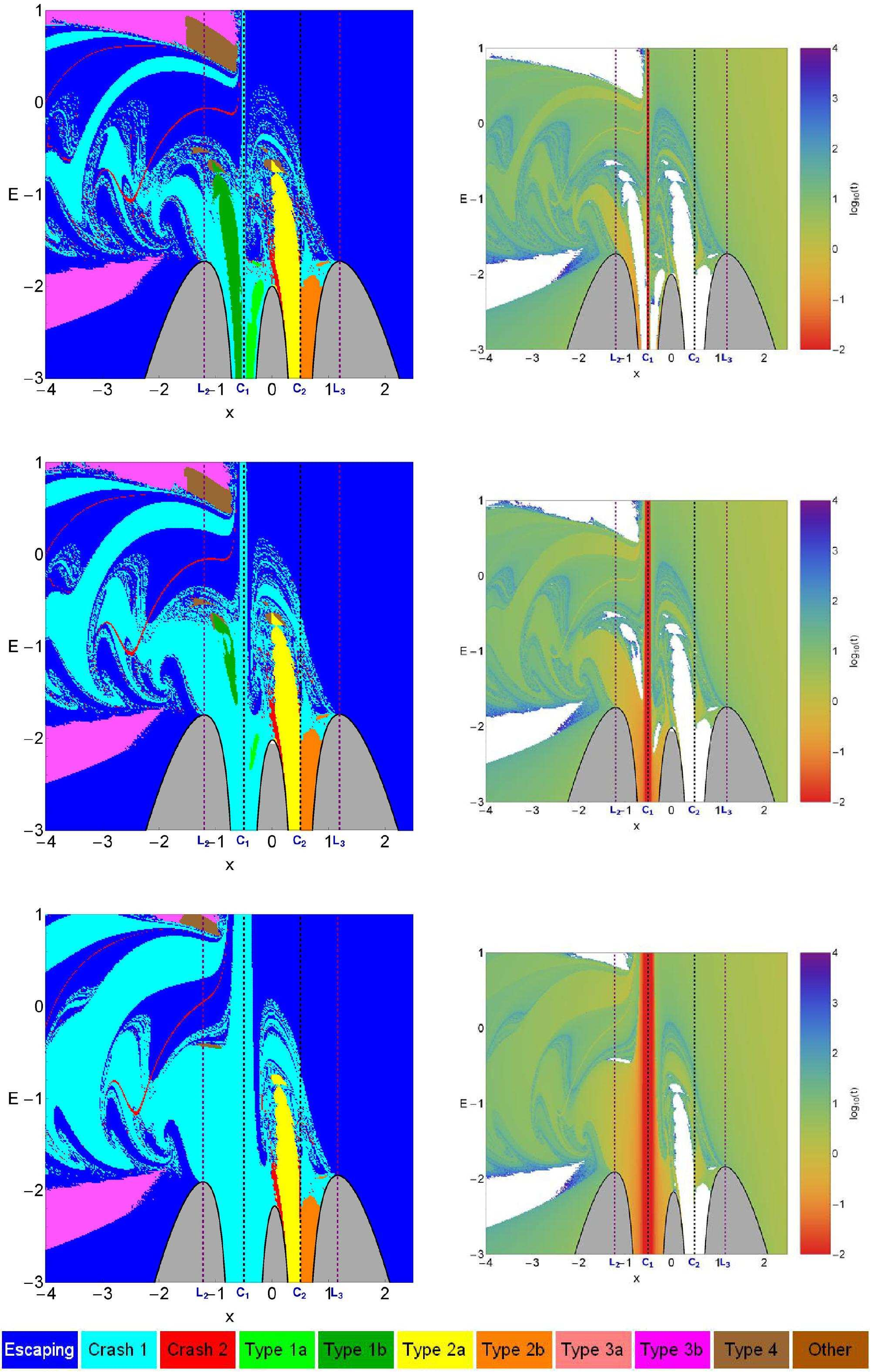}}
\caption{(left column): Orbital structure of the $(x,E)$ plane when (Top row): $A_1 = 0.001$; (middle row): $A_1 = 0.01$; (bottom row): $A_1 = 0.1$. (right column): the distribution of the corresponding escape/collisional times of the orbits. The vertical black dashed lines denote the centers of the two primaries, wile the vertical purple dashed lines indicate the position of the Lagrangian points $L_2$ and $L_3$. The color bar contains the color code which relates the types of orbits presented in Fig. \ref{types} with different colors.}
\label{xE}
\end{figure*}

\begin{figure*}[!tH]
\centering
\resizebox{0.85\hsize}{!}{\includegraphics{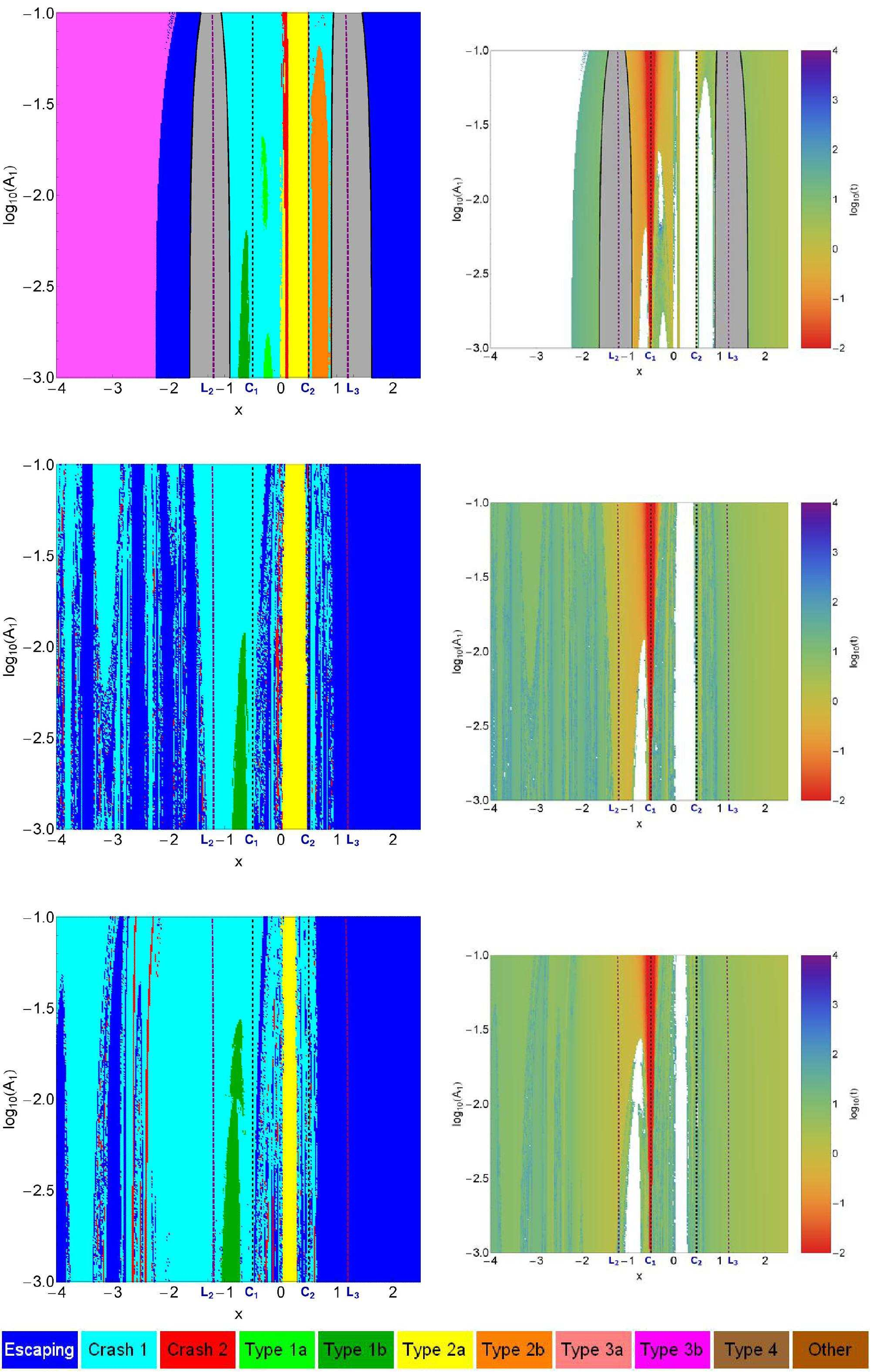}}
\caption{(left column): Orbital structure of the $(x,A_1)$ plane when (Top row): $E = -2.0$; (middle row): $E = -1.5$; (bottom row): $E = -1.0$. (right column): the distribution of the corresponding escape/collisional times of the orbits. The vertical black dashed lines denote the centers of the two primaries, wile the vertical purple dashed lines indicate the position of the Lagrangian points $L_2$ and $L_3$. The color bar contains the color code which relates the types of orbits presented in Fig. \ref{types} with different colors.}
\label{xA}
\end{figure*}

The color-coded OTDs in both parts of the physical $(x,y)$ space provide sufficient information on the phase space mixing however, for only a fixed value of the energy integral and also for orbits that traverse the surface of section either directly (progradely) or retrogradely. H\'{e}non back in the late 60s \citep{H69}, introduced a new type of plane which can provide information not only about stability and chaotic regions but also about areas of bounded and unbounded motion using the section $y = \dot{x} = 0$, $\dot{y} > 0$ (see also \citet{BBS08}). In other words, all the orbits of the test particles are launched from the $x$-axis with $x = x_0$, parallel to the $y$-axis $(y = 0)$. Consequently, in contrast to the previously discussed types of planes, only orbits with pericenters on the $x$-axis are included and therefore, the value of the energy $E$ can be used as an ordinate. In this way, we can monitor how the energy influences the overall orbital structure of our dynamical system using a continuous spectrum of energy values rather than few discrete energy levels. In the left column of Fig. \ref{xE} we present the orbital structure of the $(x,E)$ plane for three values of the oblateness coefficient when $E \in [-3,1]$, while in the right column of the same figure the distribution of the corresponding escape/collision times of orbits is depicted.

We observe the presence of several types of regular orbits around the two primary bodies. Being more precise, on both sides of the primaries we identify stability islands corresponding to both direct (counterclockwise) and retrograde (clockwise) quasi-periodic orbits. It is seen that a large portion of the exterior region, that is for $x < x(L_2)$ and $x > x(L_3)$, a large portion of the $(x,E)$ plane is covered by initial conditions of escaping orbits however, at the left-hand side of the same plane two stability islands of type 3b regular orbits are observed. Additional numerical calculations reveal that for much lower values of $x$ $(x < 5)$ these two stability islands are joined and form a crescent-like shape. Furthermore, orbits with initial conditions very close to vertical line $x = C_1$, or in other words close to the center of the oblate primary 1 collide almost immediately with it, while their portion (thickness of the line) increases for larger values of the oblateness coefficient $A_1$. We also see that crash basins to primary 1 leak outside the interior region, mainly outside $L_2$, and create complicated spiral shapes in the exterior region. On the other hand, the thin red bands represent initial conditions of orbits that collide with primary body 2. It should be pointed out that in the blow-ups of the diagram several additional very small islands of stability have been identified\footnote{An infinite number of regions of (stable) quasi-periodic (or small scale chaotic) motion is expected from classical chaos theory.}. We may say that the stability islands around primary body 2 are almost unperturbed by the shifting on the value of the oblateness. In contrast, we see that for $A_1 = 0.1$ there is no indication of regular motion around oblate primary 1, while the amount of orbits that crash into primary 1 significantly grows with increasing $A_1$. Another interesting phenomenon is the fact that as primary body 1 becomes more and more oblate the fractility of the $(x,E)$ plane reduces and the boundaries between escaping and collisional motion appear to become smoother. The following Table \ref{table1} shows the percentages of the all types of orbits in the color-coded OTDs shown in Figs. \ref{xE} and \ref{xA}. We observe that type 3a and 4 regular orbits are completely absent. It should be pointed out that the high values of escaping and crash 1 percentages are due to the extended scattering region and large oblateness, respectively.

\begin{table}[!ht]
\begin{center}
   \caption{Percentages of all types of orbits in the color-coded OTDs shown in Figs. \ref{xE}(a-c) and \ref{xA}(a-c). ((a) corresponds to top row of the figures, (b) corresponds to middle rows, while (c) corresponds to bottom rows).}
   \label{table1}
   \setlength{\tabcolsep}{2.5pt}
   \begin{tabular}{@{}lccccccccccc}
      \hline
      Figure & Escaping & Crash 1 & Crash 2 & Type 1a & Type 1b & Type 2a & Type 2b & Type 3a & Type 3b & Type 4 & Other \\
      \hline
      \ref{xE}a & 64.29 & 18.33 & 1.46 & 0.73 & 2.04 & 3.15 & 1.41 & 0.00 &  6.93 & 0.00 & 1.66 \\
      \ref{xE}b & 59.14 & 27.53 & 1.12 & 0.13 & 0.55 & 3.09 & 1.38 & 0.00 &  5.81 & 0.00 & 1.25 \\
      \ref{xE}c & 52.03 & 38.73 & 0.86 & 0.00 & 0.01 & 2.77 & 1.06 & 0.00 &  4.08 & 0.00 & 0.46 \\
      \ref{xA}a & 28.54 & 19.52 & 1.71 & 0.63 & 1.08 & 7.54 & 5.15 & 0.00 & 35.82 & 0.00 & 0.01 \\
      \ref{xA}b & 54.08 & 36.09 & 1.77 & 0.00 & 1.54 & 6.47 & 0.00 & 0.00 &  0.04 & 0.00 & 0.01 \\
      \ref{xA}c & 40.31 & 51.24 & 2.05 & 0.00 & 2.53 & 3.72 & 0.00 & 0.00 &  0.00 & 0.00 & 0.15 \\
      \hline
   \end{tabular}
\end{center}
\end{table}

In order to obtain a more complete view of the orbital structure of the system, we follow a similar numerical approach to that explained before varying now the value of the oblateness coefficient $A_1$ for three energy levels which belong to the three main Hill's regions. This allow us to construct again a two-dimensional (2D) plane in which the $x$ coordinate of orbits is the abscissa, while the logarithmic value of the oblateness coefficient $\log_{10}(A_1)$ is the ordinate. The orbital structure of the $(x,A_1)$ plane when $\log_{10}(A_1) \in [-3, -1]$ is shown in the left column of Fig. \ref{xA}, while the distribution of the corresponding escape/collision times of orbits is given in the the right column of the same figure. The black solid line is the limiting curve which distinguishes between regions of allowed and forbidden motion and is defined as
\begin{equation}
f(y,A_1) = V(x = 0,y;A_1) = E,
\label{zvc}
\end{equation}
while the vertical dashed magenta lines indicate the position of the Lagrangian points $L_2$ and $L_3$. Here it should be pointed out that as it was mentioned in Section \ref{mod} the position of the Lagrangian points $L_2$ and $L_3$ is no longer fixed with variable $A_1$, since it is function of the oblateness coefficient.

A very complicated orbital structure is reveled in the $(x,A_1)$ plane from which however, we can deduce some interesting results such as: (i) for $E = -2 < E_2$ we see a very organized structure in which all the different domains, corresponding to different types of orbits, are well-defined thus the fractility of the plane is minimum, while for larger values of the energy the boundaries of the domains become highly fractal; (ii) orbits with initial conditions very close to the center of the oblate primary 1 collide with it practically immediately; (iii) the amount of collisional orbits to primary 1 grows with increasing energy and oblateness, while that of escaping orbits seems to reduces; (iv) the extent of the area on the $(x,A_1)$ plane occupied by retrograde (clockwise rotating) orbits is reduced with increasing oblateness and is favoured only at relatively high energy levels $(E > E_4)$; (v) the stability island of retrograde motion around primary 2 is present only at low energies $(E < E_2)$ and at low values of $A_1$ $(A_1 < 0.01)$, while for larger values of energy or/and oblateness disappears; (vi) the stability island of prograde (direct) motion around primary 2 seems to be unaffected by the change on the value of the oblateness however, it slightly reduces as we proceed to higher energy levels. The phenomenon that stability islands can appear and disappear as a dynamical parameter is changed has also been reported in earlier paper (e.g., \citet{BBS06,dAT14}).

\section{Discussion and conclusions}
\label{disc}

The scope of this work was to shed some light to the properties of motion in the Copenhagen problem where one of the primaries is an oblate spheroid. We continued the work initiated in Paper I and II following similar numerical techniques therefore, this paper should be considered as Part III. We managed to distinguish between bounded, escaping and collisional orbits and we also located the basins of escape/collison, finding correlations with the corresponding escape/crash times of the orbits. Our extensive and thorough numerical investigation strongly suggests, that the overall motion of a test body under the gravitational field of two primaries is a very complicated procedure and very dependent on the value of the energy integral. To our knowledge, this is the first detailed and systematic numerical analysis on the phase space mixing of bounded motion, escape and crash in the Copenhagen problem with oblateness and this is exactly the novelty and the contribution of the current work.

We defined for several values of the total orbital energy $E$, dense uniform grids of $1024 \times 1024$ initial conditions regularly distributed on both parts ($\dot{\phi} < 0$ and $\dot{\phi} > 0$) the physical $(x,y)$ plane inside the area allowed by the value of the energy. All orbits were launched with initial conditions inside the scattering region, which in our case was a square grid with $-2\leq x,y \leq 2$. For the numerical integration of the orbits in each type of grid, we needed about between 7 hours and 8 days of CPU time on a Pentium Dual-Core 2.2 GHz PC, depending on the escape and collisional rates of orbits in each case. For each initial condition, the maximum time of the numerical integration was set to be equal to $10^4$ time units however, when a particle escaped or collided with one of the primaries the numerical integration was effectively ended and proceeded to the next available initial condition.

The present article provides quantitative information regarding the escape and crash dynamics in the Copenhagen problem with oblateness. The main numerical results of our research can be summarized as follows:
\begin{enumerate}
 \item The value of the oblateness coefficient of the primary body 1 was found to greatly influence the stability regions around the same primary body. In particular, as the primary becomes more and more oblate the size of the stability islands corresponding to regular motion around only primary 1 is reduced and at relatively high values of the oblateness there is no indication of regular motion around the oblate primary.
 \item We found that orbits with initial conditions in the vicinity of the oblate primary body 1 crash almost immediately on it, while on the other hand orbits initiated around spherical primary 2 have larger non zero crash times.
 \item In both parts of the configuration space ($\dot{\phi} < 0$ and $\dot{\phi} > 0$) the crash basins of oblate primary 1 were found to be well defined broad and extended regions, while the initial conditions leading to crash with spherical primary 2 form thin filaments and spiral bands.
\end{enumerate}

Judging by the detailed and novel outcomes we may say that our task has been successfully completed. We hope that the present numerical analysis and the corresponding results to be useful in the field of escape dynamics in the Copenhagen problem with oblateness. The outcomes as well as the conclusions of the present research are considered, as an initial effort and also as a promising step in the task of understanding the escape mechanism of orbits in this interesting version of the classical three-body problem. Taking into account that our results are encouraging, it is in our future plans to properly modify our dynamical model in order to expand our investigation into three dimensions and explore the entire six-dimensional phase thus revealing the influence of the oblateness coefficient on the orbital structure. Moreover, it would be interesting to apply our numerical methods in some interesting cases of the PCRTBP such the Earth-Moon and Saturn-Titan systems where however the values of oblateness are much smaller than those considered in the present paper.

\section*{Acknowledgments}

I would like to express my warmest thanks to J. Nagler, Ch. Jung and V. Kalantonis for all the illuminating and inspiring discussions during this research. The author would also like to thank the two anonymous referees for the careful reading of the manuscript and for all the apt suggestions and comments which allowed us to improve significantly both the quality and the clarity of the paper.

\end{document}